\documentclass[11pt,a4paper]{scrartcl}

\usepackage{CLICdp}

\usepackage{CLICdp_definitions}

\usepackage{todonotes}
\usepackage{multirow}
\usepackage{booktabs}
\usepackage{changepage}

\title{Conformal Tracking for all-silicon trackers at future electron--positron colliders}

\clicdppub{2019}{003}  

\date{\today}

\addauthor{E. Brondolin}{\institute{1}\hcomma\editor{Corresponding Author}}
\addauthor{F. Gaede}{\institute{2}}
\addauthor{D. Hynds}{\institute{1}\hcomma\thanks{Now at Nikhef, Amsterdam, The Netherlands}}
\addauthor{E. Leogrande}{\institute{1}\hcomma\editor{Corresponding Author}}
\addauthor{M. Petri\v{c}}{\institute{1}\hcomma\thanks{Now at Jozef Stefan Institute, Slovenia}}
\addauthor{A. Sailer}{\institute{1}}
\addauthor{R. Simoniello}{\institute{1}\hcomma\thanks{Now at University of Mainz, Mainz, Germany}}

\addinstitute{1}{CERN, Switzerland}
\addinstitute{2}{DESY, Germany}

\abstract{
Conformal tracking is an innovative and comprehensive pattern recognition technique
using a cellular automaton-based track finding performed in a conformally-mapped space. 
It is particularly well-suited for light-weight silicon systems with high position resolution, 
such as the next generation of tracking detectors designed for future electron--positron colliders.
The algorithm has been developed and validated with simulated data of the CLICdet tracker. 
It has demonstrated not only excellent performance in terms of tracking efficiency, fake rate
and track parameters resolution but also robustness against the high beam-induced background levels.
Thanks to its geometry-agnostic nature and its modularity,
the algorithm is very flexible and can easily be adapted to other detector designs and experimental environments
at future \epem colliders.
}

\titlecomment{This work was carried out in the framework of the CLICdp Collaboration} 

\graphicspath{ {./logos/}{./figures/} }

\newlength{\abc}
\AtBeginDocument{%
\renewcommand{\ref}[1]{\mbox{\autoref{#1}}}

}

\begin{document}

\titlepage
\newpage


\pagenumbering{arabic}

\section{Introduction}
\label{sec:introduction_CT}

Conformal tracking is a pattern recognition technique for track reconstruction that
combines the two concepts of conformal mapping~\cite{HANSROUL1988498} and cellular automata~\cite{GLAZOV1993262,KISEL200685}.
This algorithm was developed to efficiently exploit the main features of the next generation
of tracking detectors, such as their low-mass all-silicon systems and their few-\SI{}{\micron} position accuracy~\cite{Hoffman:2019meg}.
Its novel and comprehensive track finding strategy can easily be applied to cope with the beam-induced background of experiments 
at different future electron--positron colliders~\cite{cdrvol2,Benedikt:2651299,CEPCStudyGroup:2018ghi}.

This paper describes in detail the conformal tracking technique
for a generic multi-layer silicon detector in a solenoidal magnetic field
and evaluates its performance in terms of effectiveness, robustness and flexibility.
In~\cref{sec:conformal_tracking} the features of the algorithm are described.
Conformal tracking has been developed as the baseline reconstruction strategy for the 
detector CLICdet designed for CLIC, a proposed future electron--positron collider~\cite{cdrvol2, cdrupdate, clic_summaryReport}. 
The application of the algorithm to CLICdet is the focus of \cref{sec:clic_tracking}.
Firstly, the experimental conditions and physics requirements for CLIC are described.
Then, the design of the CLICdet tracking system is introduced
as well as the event simulation and reconstruction software framework. 
Finally, the tracking performance in terms of tracking efficiency and fake rate,
track parameter resolution and CPU time are presented for single isolated particles, as well as for complex event topologies.
Conclusions and next steps for the optimisation of the conformal tracking approach are discussed in~\cref{sec:summary}.

\section{Conformal Tracking}
\label{sec:conformal_tracking}

Conformal tracking algorithm can be divided into two main blocks:
the conformal mapping method and the cellular automaton-based track finding.

\subsection{Conformal mapping}
\label{sec:CT_hitmapping}

In the conformal algorithm, point coordinates in global Euclidean space $(x,y)$ are translated into the conformal space $(u,v)$~\cite{HANSROUL1988498}.
The idea behind this coordinate transformation is that circles passing through the origin of a coordinate system $(x,y)$
can be transformed into straight lines in a new coordinate system $(u,v)$.
The circle equation in global coordinates $(x,y)$
\begin{linenomath}
\begin{equation}
(x-a)^{2} + (y-b)^{2} = r^{2}
\end{equation}
is equivalent to a straight line in the $(u,v)$ plane
\begin{equation}
v = -\frac{a}{b}u + \frac{1}{2b}
\label{eq:straight_line}
\end{equation}
if the circle is passing through the origin such that $r$ is fixed to $r^2 = a^2 + b^2$
and if the following transformations are applied:
\begin{equation}
u = \frac{x}{x^{2}+y^{2}}, \qquad v = \frac{y}{x^{2}+y^{2}}.
\end{equation}
\end{linenomath}

\begin{figure}[tb]
\centering

\begin{subfigure}{0.45\textwidth}
\includegraphics[width=\linewidth]{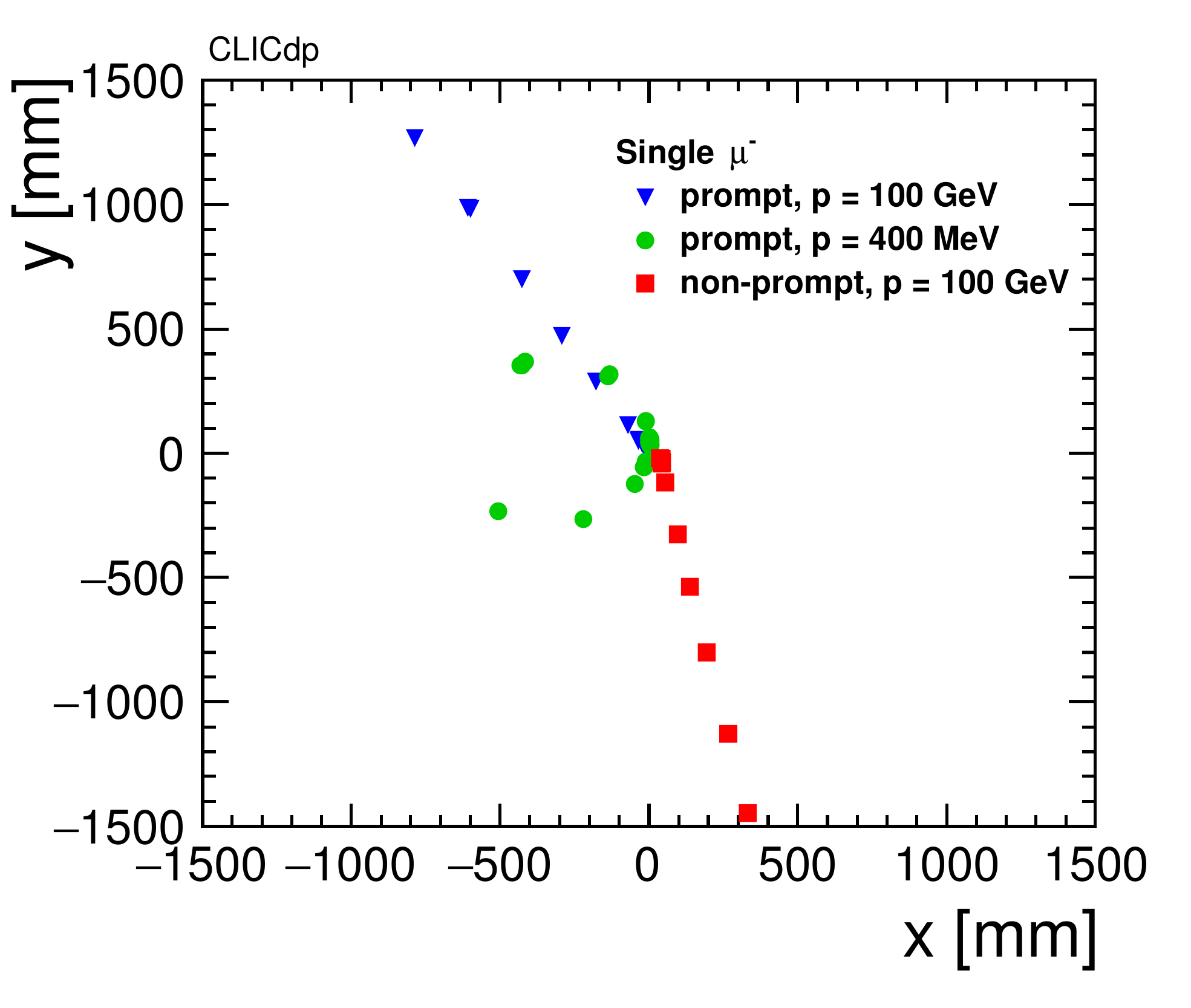}
\end{subfigure}\hfil
\begin{subfigure}{0.45\textwidth}
\includegraphics[width=\linewidth]{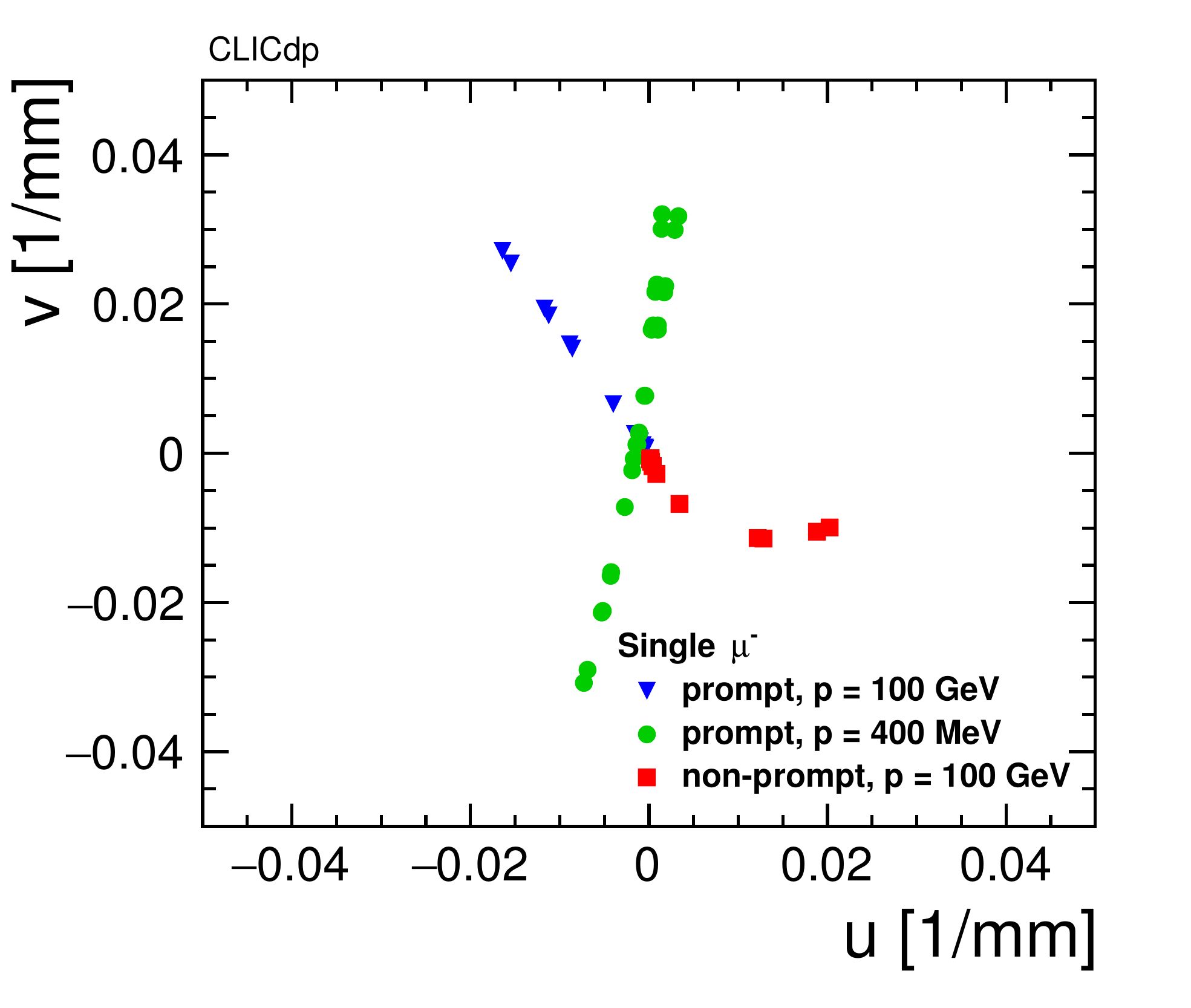}
\end{subfigure}\hfil

\caption{Examples of tracks in the $(x,y)$ global coordinate system (left) and in the $(u,v)$ conformal coordinate system (right). Hits produced by three track types are shown: prompt (blue), prompt with low momentum (green) and non-prompt (red).}
\label{fig:conformal_mapping}
\end{figure}

Through the application of the conformal mapping,
finding tracks of charged particles bent by a homogeneous magnetic field 
can be reduced to a search for straight lines.
The radial order of the hit positions is inverted in the conformal
space with respect to the global space:
hits on the innermost part of the detector are mapped to outer regions in the $(u,v)$ plane and vice versa.

The simple pattern recognition technique suggested in original applications of conformal mapping
consists of grouping hits aligned in the same direction in the $(u,v)$ 
plane, by searching for peaks in the angular hit distribution in conformal space~\cite{HANSROUL1988498}.
However, this method does not take into account deviations from the straight line path, which can arise in real measurements.
These deviations can be either a result of \emph{a priori} deviations from the circular path in global space, such
as for particles undergoing multiple scattering, or due to mathematical approximations introduced in the conformal mapping formulas, as is 
the case of particles not produced at the origin of the $(x,y)$ plane, also known as \emph{displaced} or \emph{non-prompt} particles.

This is illustrated in~\cref{fig:conformal_mapping}, where hits belonging to three muon tracks simulated
in the CLICdet tracker, described in~\cref{sec:clic_tracker}, are shown in both global and conformal space.
One track is produced by a prompt high-momentum muon passing through the origin of the
$(x,y)$ plane and unperturbed in its path, the second one corresponds to a prompt low-momentum muon undergoing multiple scattering, 
and the last one is a track produced by a non-prompt muon.
In the latter two cases, the deviation of the trajectories from straight lines is clearly visible.
To take these deviations systematically into account, pattern recognition in conformal space is performed via cellular automaton (CA).

\subsection{Cellular tracks reconstruction}
\label{sec:CA}

The CA-based algorithm~\cite{KISEL200685} used for track finding in conformal tracking consists of the creation and extension of cells, 
which are defined as segments connecting two hits.
It is used to reconstruct \emph{cellular tracks} and consists of two main algorithms:
the building and the extension of cellular track candidates.
These two algorithms can be used in a recursive way as the final track finding strategy, as shown in~\cref{subsubsec:chain}
for the case of the CLICdet detector.

\subsubsection{Building of cellular track candidates}
\label{sec:CA_build}

Since the CA can potentially create a vast number of cells, the first step is to create a small number of \emph{seed} cells, which provide the basic set of track candidates to be further extended. 
This is done by limiting the seeding procedure only to a subset of hits, hereafter referred to as \emph{seeding collection}\footnote{The choice of the subset of hits for the seeding collection is flexible and can be conveniently adapted to each use case. The specific choice for CLIC is summarised in \cref{tab:iterations}.}.

\paragraph{Seed cell creation} 

As a first step, hits in the seeding collection are sorted in decreasing conformal radius 
and each individual hit of the seeding collection is considered as a \emph{seed hit}. 
From each seed hit, a search for neighbouring hits is performed in polar angle\footnote{Defined as $\Theta = \arctan(u/v) + \pi$.}: 
hits with a polar angle within the search window 
$[\Theta_{\textnormal{seed hit}}-\Delta\Theta_{\textnormal{neighbours}},\Theta_{\textnormal{seed hit}}+\Delta\Theta_{\textnormal{neighbours}}]$ 
are considered, where $\Delta\Theta_{\textnormal{neighbours}}$ is a fixed parameter.
To speed up the process, the hits are organised in k-d trees, which are 
binary tree structures optimised for a fast nearest-neighbour search~\cite{Kennel:2004}.
One or more \emph{seed cells} are created by connecting each seed hit with its nearest neighbours.
Seed cells constitute the basic structures from which the final tracks will evolve.
Seed cells are created, as long as the neighbours
do not lie on the same detector layer as the seed hit, are located at a smaller conformal radius and are not already part of a track. 
At this stage, seed cells can be discarded if their length exceeds a maximum cell length $l_{\textnormal{max}}$.

The seed cell creation plays an important role in the performance of the CA. Both the search window amplitude $\Delta\Theta_{\textnormal{neighbours}}$ and the maximum cell length $l_{\textnormal{max}}$ will have a direct effect on the efficiency of the track finding, while the number of cells produced will influence the processing time. Therefore, these parameters are tuned according to the reconstruction step and to the detector specifics. The binary search tree suffers from the limitation that the polar angle is restricted to $[-\pi,\pi]$, thus hits at either extreme are not considered as neighbours. This has been solved by extending the tree with entries occurring close to the boundaries entered twice, with polar angle values differing by $2\pi$.

\paragraph{Seed cell extrapolation} 

Once all seed cells are created from one seed hit, they are extrapolated along the cell direction for a given distance, before a new nearest-neighbour search is performed from the extrapolated points. 
A different k-d tree is used, which allows for a search for neighbouring hits in magnitude of conformal radius, i.e. within the search window $[R_{\textnormal{seed hit}}-\Delta R_{\textnormal{neighbours}},R_{\textnormal{seed hit}}+\Delta R_{\textnormal{neighbours}}]$, 
where $\Delta R_{\textnormal{neighbours}}$ is a fixed parameter.
This is a more conservative approach than the search in polar angle, as it allows for cell extension 
without any assumption on $\Delta\Theta_{\textnormal{neighbours}}$. 
All the neighbouring hits found in the search window 
are used to create a cell connected to the seed cell, provided that they do not lie on the same detector layer as the endpoint of the seed cell,
have smaller conformal radius and have not been already used in a track. 
Newly-created cells which form an angle larger than $\alpha_{\textnormal{max}}$ with the seed cell are discarded at this point. 
Similarly to the parameters $\Delta\Theta_{\textnormal{neighbours}}$ and $l_{\textnormal{max}}$, also $\Delta R_{\textnormal{neighbours}}$ and $\alpha_{\textnormal{max}}$ are parameters to be tuned according to the reconstruction step and detector specifics. 
The remaining cells are further extrapolated in the same manner as the seed cells.

\paragraph{Cellular track candidates} 

Each cell contains information about its start and end points as well as a \emph{weight}.
The weight of a cell indicates how many other cells are further connected to it.
Each subsequent link increments the cell weight by one unit, such that the higher the weight,
the higher the potential of the cell to make a track.
\emph{Cellular track candidates} are chains of cells, created by all cells compatible within a certain
angle window starting from the highest weighted seed cell. A minimum number of hits for each candidate is required ($N^{\textnormal{hits}}_{\textnormal{min}}$).
Cellular track candidates are then fitted with a linear regression to obtain the parameters in \cref{eq:straight_line}. 

One immediate impact of performing pattern recognition in the 2-dimensional conformal space is the loss of information due to the projection from global 3-dimensional space. To recover this information, an additional fit is performed in the $(s,z)$ plane, where $s$ is the coordinate along the helix arc segment. As $s$ varies linearly with the track length along the global $z$ axis, a linear regression is applied here as well.
The $\chi^{2}_{(u,v)}$ and $\chi^{2}_{(z,s)}$ from both linear regressions are calculated and normalised by the number of degrees of freedom. Only the cellular tracks for which neither of the two normalised $\chi^{2}$ values exceeds a threshold $\chi^{2}_{\textnormal{max}}$ are considered as valid track candidates. This threshold is chosen according to the reconstruction step and detector specifics. A further attempt to recover good cellular tracks containing a spurious hit is made by removing, one by one, each hit on the track, refitting and recomputing the normalised $\chi^{2}_{(u,v)}$ and $\chi^{2}_{(z,s)}$. 
This procedure does not prevent tracks which share hits from being accepted as valid track candidates. Therefore, for all tracks created by a single seed hit, \emph{clones} are defined as tracks sharing more than one hit and are filtered according to a combined length and $\chi^{2}_{tot}$ criterion. The $\chi^{2}_{tot}$ is calculated as the sum of the squares of $\chi^{2}_{(u,v)}$ and $\chi^{2}_{(z,s)}$. In general, longer tracks are preferred, while the one with the best $\chi^{2}_{tot}$ is chosen in case of equal length.

The building of cellular track candidates stemming from the same seed hit as described above is repeated for each seed hit.
At the end of the process, all hits belonging to the created cellular tracks are marked as \emph{used}.

\subsubsection{Extension of cellular track candidates}
\label{sec:CA_extend}

The cellular tracks that are the output of the \emph{building} procedure described above, are extended with the \emph{extension} algorithm. 
The latter requires as input the hit collection with which to perform the cellular track extension.
For each cellular track candidate an estimation of the particle transverse momentum is obtained 
using the parameters extracted from the linear regression fit in the $(u,v)$ space.
Slightly different approaches are followed for track candidates with an estimated transverse momentum above or below a given \pT threshold ($p_{\textnormal{T, cut}}$). The latter can be tuned according to the reconstruction step and detector specifics.
Tracks with \pT higher than the threshold are extended first, as they are easier to reconstruct. Consequently, a lower number of hits is left for the extension of the lower-\pT tracks.

The track extension for higher-\pT tracks proceeds in a similar manner as described above as far as the CA is concerned. 
The endpoints of the previously formed cellular tracks are used as seed hits and a search for nearest neighbours in polar angle is performed, although limited to hits in the adjacent detector layer. 
New cellular tracks are created as the track candidates are extended to every valid neighbour, among which the best track candidate is chosen based on the smallest $\chi^{2}_{tot}$ normalised by the number of degrees of freedom. Such extension of the track candidate with the best hit per layer is repeated until the last detector layer. Further details, in the context of track reconstruction in the CLICdet tracker, are given in~\cref{subsubsec:chain}.

It is important to note that, although the track extension relies on the layer order, no condition is imposed on the position and number of layers. In this respect, the algorithm is transparent to the specific detector geometry. Moreover, if no best hit is found on one layer, the CA proceeds on the subsequent layer, avoiding loss of track finding efficiency.

Once the reconstruction of higher-\pT tracks is completed and their hits are marked as used, 
the extension procedure is performed with the remaining hits for finding lower-\pT tracks, such as particles looping in the tracker.
Since their trajectories deviate from the straight lines in conformal space due to multiple scattering, as shown in \cref{fig:conformal_mapping}, no nearest neighbour search is applied at this point. All unused hits are considered for track extension, provided that they are not located on the other side of the detector in $z$ with respect to the seed hit. The CA is performed in the same way as for the building of cellular track candidates, except for a quadratic term introduced in the regression formula, in order to take into account deviations from a straight line~\cite{HANSROUL1988498}.

\section{Track reconstruction at CLIC}
\label{sec:clic_tracking}

The effectiveness and robustness of the conformal tracking algorithm is presented using as an example the track reconstruction in the CLICdet tracker, where conformal tracking is adopted as the track finding technique.

\subsection{Experimental conditions and tracker requirements}
\label{sec:intro_CLIC}

CLIC is a proposed \epem collider operating in three stages at 
centre-of-mass energies of \SI{380}{\GeV}, \SI{1.5}{\TeV} and \SI{3}{\TeV}~\cite{cdrvol2, cdrupdate, clic_summaryReport}. 
Its main goals are to measure the properties of the top quark and the Higgs boson with high precision 
and to search for physics beyond the Standard Model.

The scientific goals of CLIC place demanding requirements on the performance of the tracking system 
used in the CLIC detector (CLICdet)~\cite{cdrvol2,Hoffman:2019meg}:
\begin{itemize}
\item excellent transverse momentum resolution for high-momentum tracks in the central detector region, at the level of $\sigma_{\pT}/\pT^2 \leq \SI{2e-5}{\per\GeV}$;
\item precise impact-parameter resolution, at the level of $\sigma_{d_0}^2 = (\SI{5}{\um})^2 + (\SI{15}{\um\GeV})^2/(p^2\sin^{3}\theta)$, 
to allow for accurate vertex reconstruction, thus enabling precise $\PQb$-, $\PQc$-, and light-quark jet tagging.
\end{itemize}

Another main challenge for successfully achieving the physics goals is the 
reconstruction and identification of the \gghadron{} events
produced by the beamstrahlung emitted by the 
electron and positron bunches traversing the high field of the opposite beam.
These events produce particles with high energy at large angles with respect to the beam line and thus can affect the performance of the reconstruction algorithm.

To meet the requirements dictated by its ambitious physics programme
and to cope with its challenging beam-driven background levels, the CLICdet features an ultra-light
silicon tracking system with precise hit-timing and hit-position resolutions~\cite{Hoffman:2019meg}.

\subsection{The CLICdet tracker}
\label{sec:clic_tracker}

For the CLIC detector studies, a right-handed coordinate system is in use, 
with the origin at the centre of the detector, the \emph{z} axis along the detector 
axis close to the beam direction and the \emph{y} axis pointing upwards with respect to the ground. 
The polar angle $\theta$ is defined relative to the positive \emph{z} axis and 
the azimuthal angle $\phi$ is defined relative to the \emph{x} axis in the (\emph{x,y}) plane. 

\begin{figure}[tb]
  \centering
  \begin{subfigure}[!b]{0.48\textwidth}
    \includegraphics[width=\textwidth]{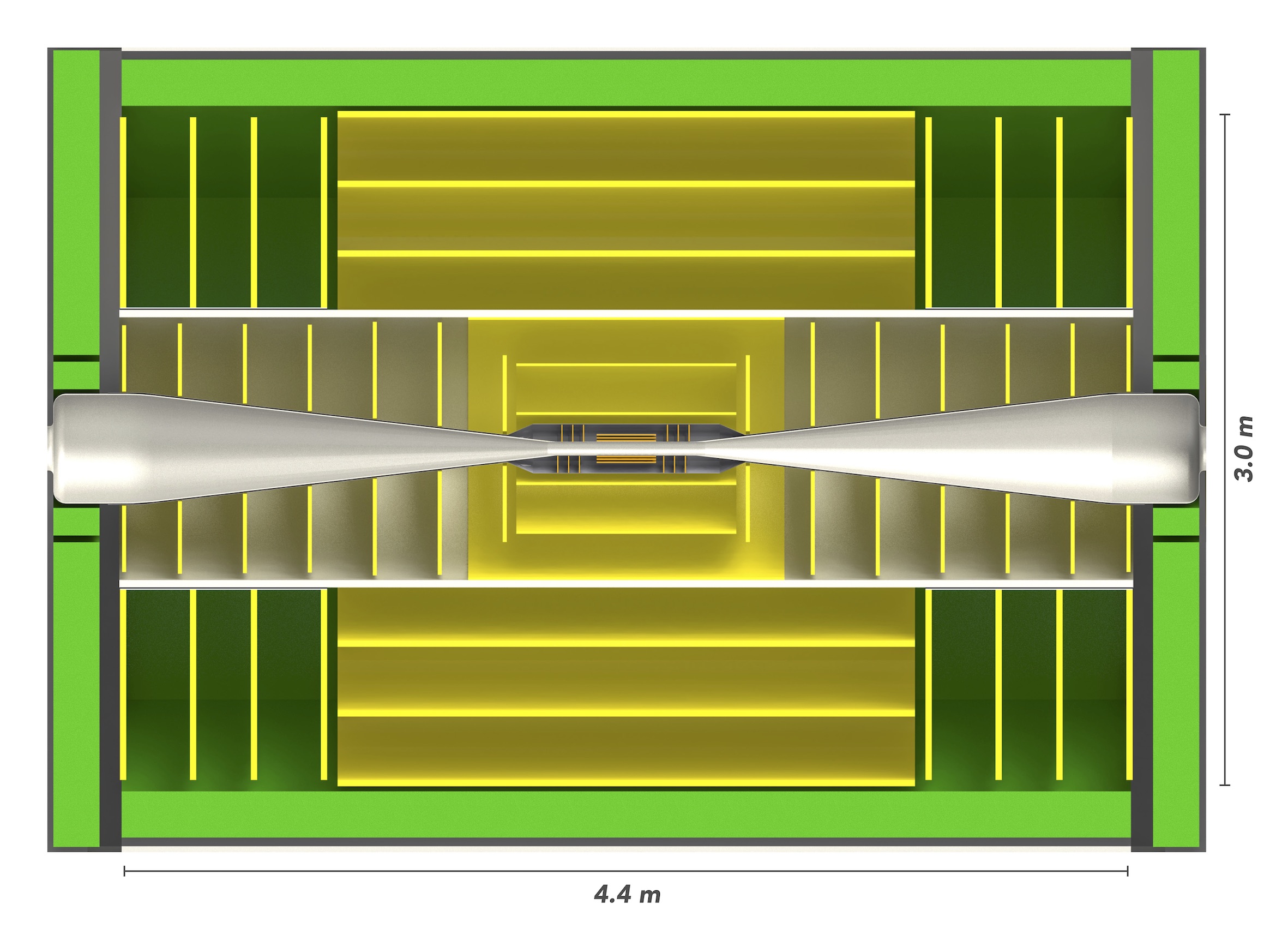}
  \end{subfigure}
  \hfill
  \begin{subfigure}[!b]{0.48\textwidth}
    \includegraphics[width=\textwidth]{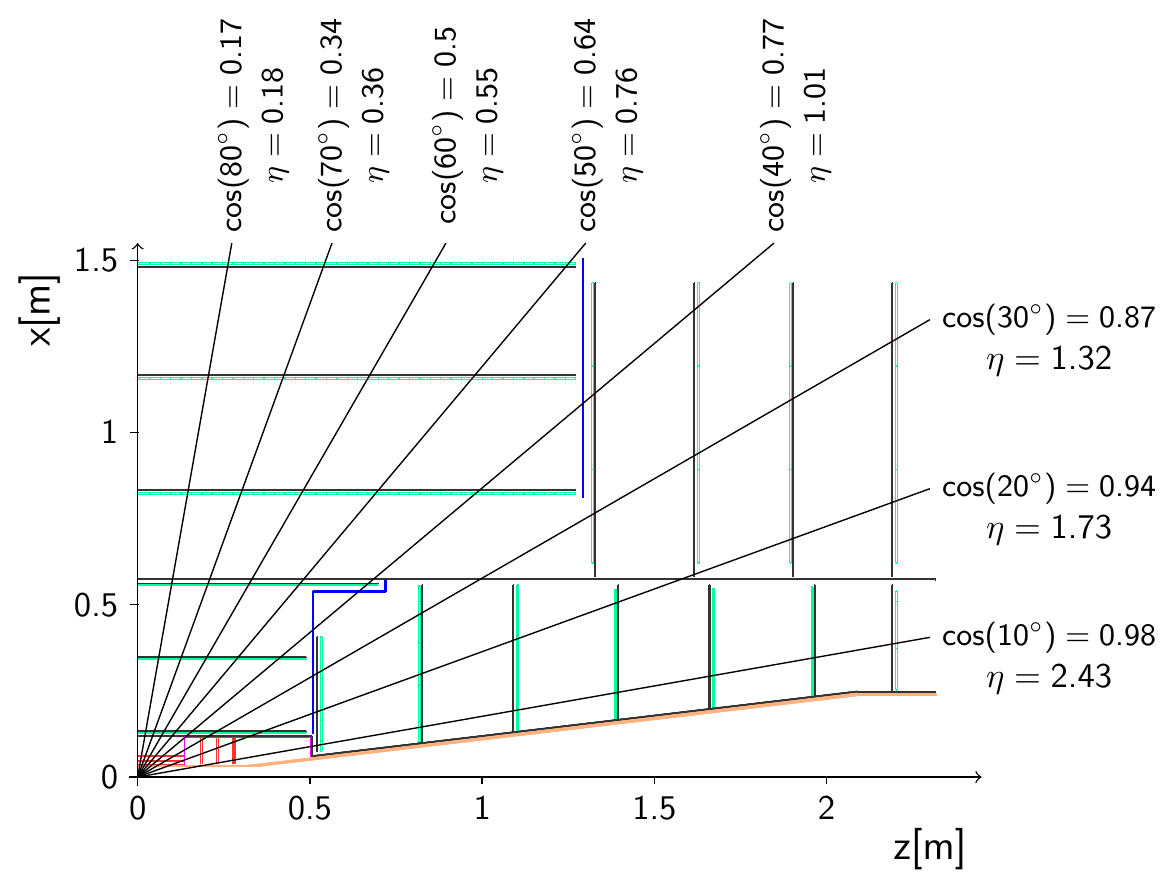}
  \end{subfigure}
  \caption{(left) Cut view of the tracker in CLICdet~\cite{CLICdet,CLICperf}: the vertex-detector layers are shown in orange, while
  the main tracker layers are depicted in yellow. The tracker is surrounded by the electromagnetic calorimeter, depicted in bright green.
  (right) A quarter of the tracking system layout in the $(x,z)$ plane~\cite{CLICdet}.}
  \label{fig:CLICdet}
\end{figure}

The CLICdet layout follows the typical collider detector scheme, and
its design is optimised for the \SI{3}{\TeV} CLIC energy stage.
Its tracking device, placed in the innermost part of CLICdet,
is an all-silicon system composed of a vertex detector and a large tracker volume, immersed in a magnetic field of \SI{4}{\tesla} produced by a superconducting solenoid.
Its layout is shown in~\cref{fig:CLICdet} (left)~\cite{CLICdet,CLICperf}.

The vertex detector is made of $25\times25~\si{\micron}^2$ pixels.
They are arranged in three cylindrical barrel double-layers 
at radii of $31-$\SI{33}{\mm}, $44-$\SI{46}{\mm} and $58-$\SI{60}{\mm} and with half length of \SI{130}{\mm}.
In the forward region, the pixel modules are organised in flat trapezoidal double-layers,
also called petals, placed in a spiraling arrangement to allow for a turbolent air flow
for efficient cooling of the vertex detectors.
There are 24 petals in total per side with a radius ranging from \SI{33}{\mm} to \SI{102}{\mm}. 
The first petal is located at $|z| = \SI{160.0}{\mm}$ from the interaction point and the last petal at $|z| = \SI{298.8}{\mm}$.
The vertex detector is an extremely accurate subdetector with a single point resolution of $3~\micron$. 
Its material budget amounts to only $0.2\%$ radiation length ($X_0$) per single layer.

The main tracking detector is made of silicon micro-strips
and is divided by a light-weight support tube into two regions: the Inner Tracker and the Outer Tracker.
In the Inner Tracker, the silicon strips are arranged in three barrel layers and seven forward disks per side,
while in the Outer Tracker the silicon strips are arranged in three barrel layers and four disks per side in the forward region.
The main parameters for the tracker layers are summarised in~\cref{tab:tracker_params}.
The innermost disk in the Inner Tracker is pixelated similarly to the vertex layers, while
all other layers are made of micro-strips with larger dimensions (see ~\cref{tab:tracker_params}).
The whole tracker volume has a radius of \SI{1.5}{\meter} and a half-length of \SI{2.2}{\meter}
with a total material budget smaller than $10\% X_0$ in the most central barrel region.

\begin{table}
\caption{The main parameters of the tracker barrel layers and forward disks. In the case of the barrel layers, the inner radius ($R$) and half-length ($L/2$) are specified,
while for the forward disks the absolute $z$-position and radius range ($R$) are listed.}
\label{tab:tracker_params}
\centering
\begin{tabular}{l l c c c}
    \toprule
    Description & Name & $R$ [mm] & $L/2$ [mm] & $|z|$ [mm]\\
    \midrule
    Layer 1 & ITB1 &127 & 482 &  \\    
    Layer 2 & ITB2& 340& 482  & \\
    Layer 3 & ITB3& 554& 692 & \\
    \midrule
    Layer 4 & OTB1& 819& 1264  & \\
    Layer 5 & OTB2& 1153 &  1264 &     \\
    Layer 6 & OTB3& 1486& 1264 &\\
    \midrule
    Disk 1 & ITD1 & $72-404$  & & 524  \\
    Disk 2 & ITD2 & $99-552$  & & 808  \\
    Disk 3 & ITD3 & $131-555$ & & 1093 \\
    Disk 4 & ITD4 & $164-542$ & & 1377 \\
    Disk 5 & ITD5 & $197-544$ & & 1661 \\
    Disk 6 & ITD6 & $231-548$ & & 1946 \\
    Disk 7 & ITD7 & $250-537$ & & 2190 \\
    \midrule
    Disk 8  & OTD1 & $618-1430$ & & 1310  \\
    Disk 9  & OTD2 & $618-1430$ & & 1617  \\
    Disk 10 & OTD3 & $618-1430$ & & 1883  \\
    Disk 11 & OTD4 & $618-1430$ & & 2190  \\
    \bottomrule
\end{tabular}
\end{table}

The layout of the CLICdet tracker as implemented in the simulation model is shown in~\cref{fig:CLICdet} (right)
including the support material, cables and cooling.
All envisaged silicon sensors have a thickness of \SI{200}{\micron} including electronics.
They are assembled in modules of either $15\times$\SI{15}{mm\squared} or $30\times$\SI{30}{mm\squared}
in the simulation model.
In~\cref{tab:detSize} the expected single point resolution and sizes 
for all pixels and strips both in the vertex and in the tracker are reported.
Note that the cell sizes are driven by occupancy studies~\cite{Nurnberg_Dannheim_2017}, while the resolutions
are the values currently used in the reconstruction software.

\begin{figure}[tb]
  \centering
  \begin{subfigure}[!b]{0.48\textwidth}
    \includegraphics[width=\textwidth]{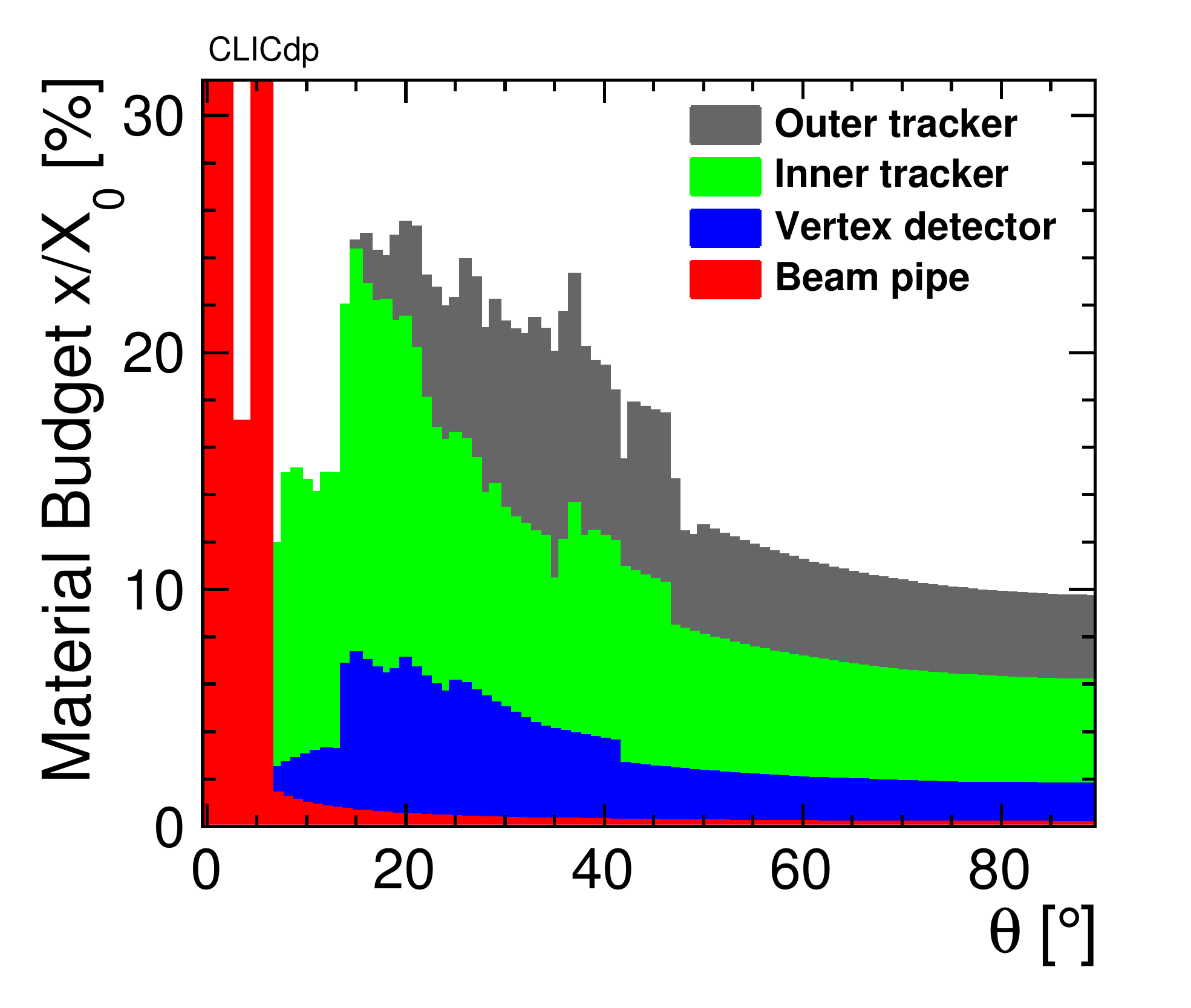}
  \end{subfigure}
  \hfill
  \begin{subfigure}[!b]{0.48\textwidth}
    \includegraphics[width=\textwidth]{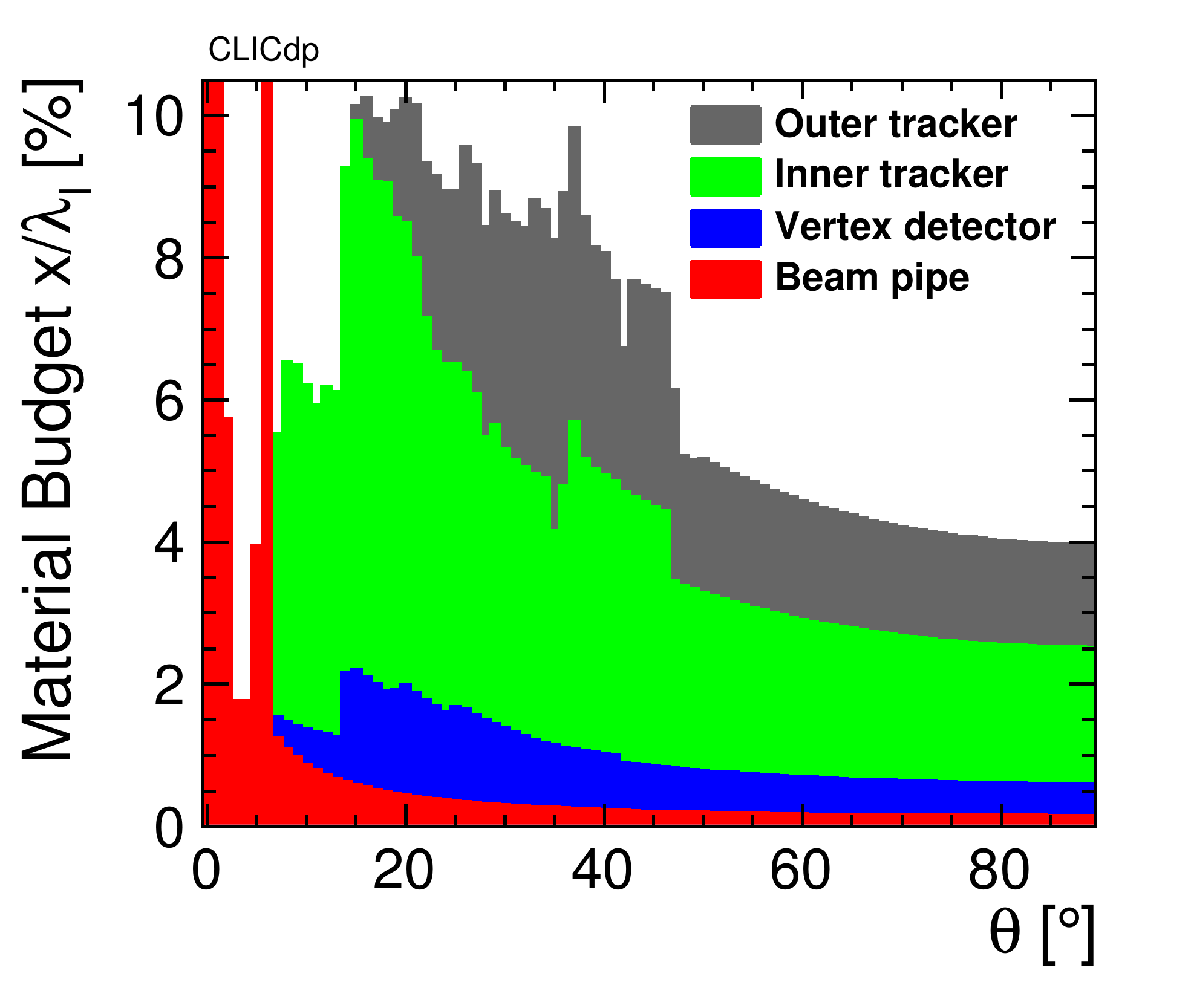}
  \end{subfigure}
  \caption{Total thickness $x$ of the tracker material
  as a function of the polar angle and averaged over azimuthal angles, expressed in units of radiation length
  $X_0$ (left) and nuclear interaction length $\lambda_I$ (right). The contribution to the total material budget 
  of different regions inside the tracking system, including sensitive
  layers, cables, supports and cooling, is shown~\cite{CLICperf}.}
  \label{fig:CLICdet_matbug}
\end{figure}

The total amount of material expressed both in terms of radiation length
$X_0$ and nuclear interaction length $\lambda_I$ is shown in~\cref{fig:CLICdet_matbug}.
All silicon tracking elements have a single hit-time resolution of $10/\sqrt{12}$~ns $\simeq 3$~ns.

\begin{table}[bt]
  \centering
  \caption{Pixel and strip sizes and corresponding single point resolutions as assumed in the simulation
  for vertex and tracking detectors. The smaller numbers for each of the strips refer to the direction 
  in the bending plane~\cite{CLICperf}. Note that the cell sizes are driven by occupancy studies~\cite{Nurnberg_Dannheim_2017}, while the resolutions
are the values currently used in the reconstruction software.}
  \label{tab:detSize}
  \begin{tabular}{l *2{l@{$\:\times\,$}r}} \toprule
  Subdetector	            & \tabtt{Cell sizes} & \tabtt{Resolutions}                                   \\ \midrule
  Vertex (Barrel and Disks) & \SI{25}{\micron}     & \SI{25}{\micron} & \SI{3}{\micron} & \SI{3}{\micron}  \\
  Inner Tracker Disk 1      & \SI{25}{\micron}     & \SI{25}{\micron} & \SI{5}{\micron} & \SI{5}{\micron}  \\
  Inner Tracker Disks 2--7  & \SI{50}{\micron}     & \SI{1}{mm}       & \SI{7}{\micron} & \SI{90}{\micron} \\
  Outer Tracker Disks       & \SI{50}{\micron}     & \SI{10}{mm}      & \SI{7}{\micron} & \SI{90}{\micron} \\
  Inner Tracker Barrel 1--2 & \SI{50}{\micron}     & \SI{1}{mm}       & \SI{7}{\micron} & \SI{90}{\micron} \\
  Inner Tracker Barrel 3    & \SI{50}{\micron}     & \SI{5}{mm}       & \SI{7}{\micron} & \SI{90}{\micron} \\
  Outer Tracker Barrel 1--3 & \SI{50}{\micron}     & \SI{10}{mm}      & \SI{7}{\micron} & \SI{90}{\micron} \\ \bottomrule
  \end{tabular}
\end{table}

\subsection{Event simulation}
\label{sec:simulation}

The CLICdet tracking performance presented in this paper is obtained with full simulation studies. 
The CLICdet detector geometry is described with the DD4hep software~\cite{Frank_2014,frank_markus_2018_1464634,frank15:ddg4} 
and simulated in \geant~\cite{Agostinelli2003, Allison2006, Allison2016186}. 
A homogeneous magnetic field of \SI{4}{T} is implemented in the simulation.
The conformal tracking software is implemented in the linear collider Marlin framework~\cite{MarlinLCCD}\footnote{Work is ongoing to translate the implementation in the GAUDI software framework~\cite{Barrand:2001ny}.}.
The tracking algorithm makes use of specific geometry information useful for the reconstruction provided by DD4hep~\cite{sailer17:ddrec}.

\begin{figure}[htb]
\centering
\includegraphics[width=0.45\linewidth]{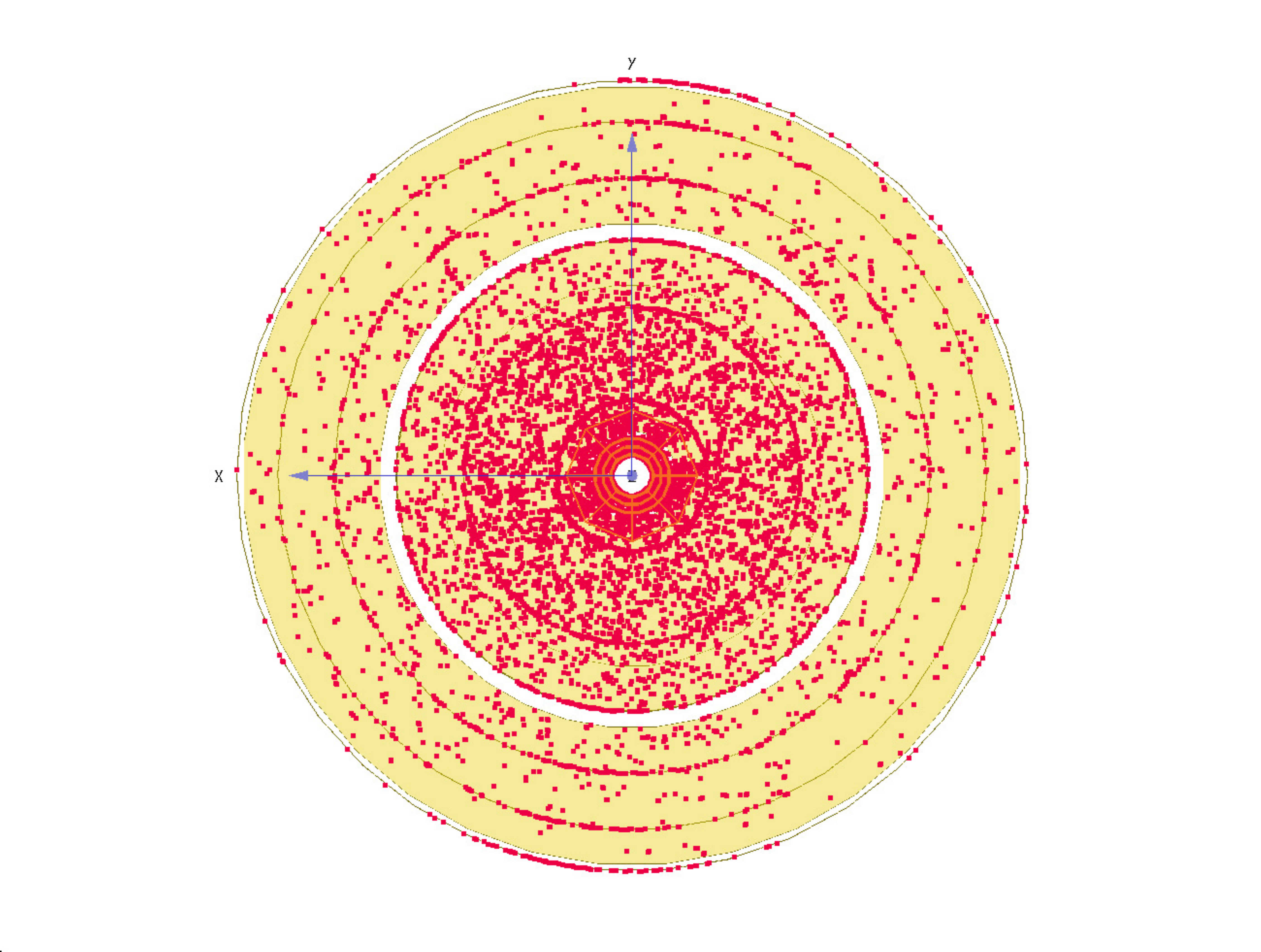}
\includegraphics[width=0.5\linewidth]{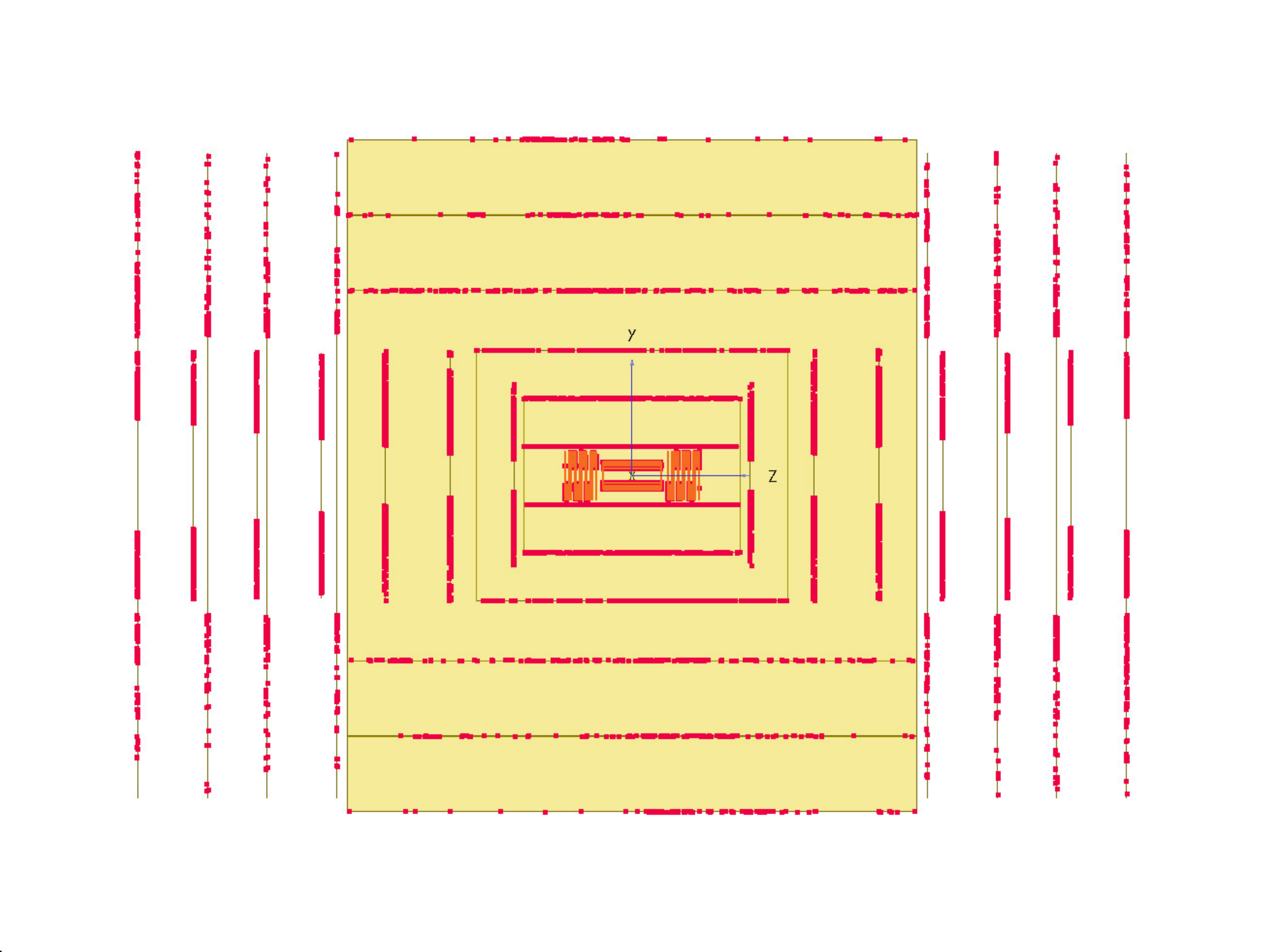}
\caption{Reconstructed hits produced in the CLICdet tracking system by a \SI{3}{\TeV} centre-of-mass energy $\epem\rightarrow\ttbar$ event simulated with 30 bunch crossings of \gghad background overlaid
shown in the $x-y$ (left) as well as in the $y-z$ (right) plane.}
\label{fig:hitsDisplay}
\end{figure}

The performance of the conformal tracking algorithm is investigated by simulating single isolated particles as well as realistic topologies in a high occupancy environment:
\begin{itemize}
\item
The isolated particle samples used in these studies consist of single muons, electrons or pions, generated with \geant. However, as a result of the simulation of the propagation of these particles through the detector, secondary particles may be present due to the interaction with the material. According to the specific case, the isolated primary particles have been generated either with fixed \pT and flat distribution in polar angle inside the tracker acceptance or with fixed polar angle and energy distribution of the type $1/E$, with \SI{100}{MeV} < $E$ < \SI{500}{GeV}.
\item
The realistic topologies examined are \epem$\rightarrow$ \ttbar events with \SI{3}{\TeV} and \SI{380}{\GeV} centre-of-mass energy, generated with \whizard\cite{Kilian:2007gr} and simulated through the detector. 
The reconstruction of the same samples is performed also with the overlay of beam-induced background events, generated with \whizard for \SI{3}{\TeV} and \SI{380}{\GeV} CLIC conditions, and superimposed on the signal process~\cite{LCD:overlay}. 
Occupancy studies for the main backgrounds in CLIC are reported in~\cite{Arominski:2668425}.
The main source of background considered in these studies are \gghadron events (cf.~\cref{sec:intro_CLIC}). 
On average, 3.2 \gghadron events occur per bunch crossing at a centre-of-mass energy of \SI{3}{\TeV}, while of the order of one bunch crossing per train contains a \emph{hard interaction}. 
The CLIC beams at the \SI{3}{\TeV} energy stage are colliding in bunch trains at a rate of 50~Hz. Each train consists of 312 bunches separated by \SI{0.5}{\ns} from each other.
In these studies, a \SI{15}{\ns} time window of background around each \epem$\rightarrow$ \ttbar event is integrated, corresponding to 10 bunch crossings before\footnote{The 10 bunch crossings preceding the physics event are integrated to include backscattered particles and shower tails.} and 20 bunch crossings after the physics event. An event display of a \SI{3}{\TeV} centre-of-mass energy \ttbar event simulated in CLICdet with 30 bunch crossings of \gghad background overlaid is shown in ~\cref{fig:hitsDisplay}.
\end{itemize}

Large simulation and reconstruction samples were produced with the iLCDirac grid production system~\cite{ilcdirac13}.

\subsection{Track reconstruction}
\label{sec:track_reco_clic}

Track reconstruction refers to the process of using a sequence of algorithms to estimate the momentum and position parameters of
a charged particle crossing the detector.
The inputs to the sequence are the reconstructed hits described in~\cref{sec:local_reconstruction}, 
while the final outputs are tracks which retain not only the hits they are made of, but also the information of the estimated track parameters.
The number of parameters that describe a track can be reduced to five~\cite{fruehwirth2000data}: two parameters for the impact
point positions, two describing the direction of the track at the impact point and one for the particle momentum.

The process of track reconstruction can be summarised as a sequence of four main algorithms:
\begin{itemize}
\item local-to-global hit coordinate transformation;
\item pattern recognition or track finding;
\item track fitting;
\item track selection.
\end{itemize}

The first algorithm is a standard coordinate transformation from the local system of the detector layer to the global $(x,y,z)$ system, thus it will not be discussed here in any further detail.
In the second algorithm, hits are grouped into track candidates. Many different track finding techniques exist, 
which have been used in past and present high-energy physics experiments~\cite{Strandlie:2010zz}. 
The strategy adopted for CLICdet is based on the conformal tracking presented in~\cref{sec:conformal_tracking}.
The third algorithm allows trajectories to be associated with the previously found track candidates. 
The results of this algorithm are for every track candidate the estimated parameters of a helix, 
their covariance or error matrix and a $\chi^{2}$ statistic.
The fitting technique applied for CLICdet is based on a Kalman filter and smoother~\cite{Fruehwirth:KF}.    
The last algorithm is optional and may target specific issues or needs, such as the reduction of spurious tracks or clones.

The sequence of algorithms and the methodology as presented below reflects the status of the software released on 09th July 2019~\cite{hynds_daniel_2019_2708196}.

\subsubsection{Hit treatment}
\label{sec:local_reconstruction}

As a first step of the track reconstruction process the \emph{simulated hits} created
by charged particles crossing the detector sensitive layers are transformed into \emph{reconstructed hits}.
The reconstructed hit positions and their uncertainties are defined 
in a local orthogonal coordinate system in the plane of each sensor. 
To obtain the reconstructed hit positions, the local positions of all simulated hits
are smeared with Gaussian distributions with width equal to the assumed single point resolution of the
subdetector in each coordinate.
The corresponding position uncertainties are then set as the single point resolution of the subdetector.
This procedure is done for all simulated hits left by the Monte Carlo particles
produced both in the physics and the background events.
Additional information is transferred from the simulated hits
to the corresponding reconstructed hits. 
This includes for example the energy deposition as well as the time of arrival\footnote{Zero detector dead time is assumed.}. 
The new collection of produced hits is used as input for pattern recognition.

\subsubsection{Track finding}
\label{subsubsec:chain}

Conformal tracking, as described in~\cref{sec:conformal_tracking}, is used as pattern recognition algorithm for track reconstruction in CLICdet.
Firstly, the conformal mapping is used to translate the hit positions from global space $(x,y)$ into the conformal space $(u,v)$.
Then, the two CA algorithms of \emph{building} (\cref{sec:CA_build}) and \emph{extension} (\cref{sec:CA_extend})
are run recursively on different hit collections and with an adapted sets of parameters to reconstruct cellular tracks.
A schematic overview of the full pattern recognition chain is summarised in~\cref{tab:iterations}, 
including the values of the parameters used in the CA-based track finding, optimized for \ttbar events at \SI{3}{TeV}.
For all steps, the window size of the nearest neighbour search in polar angle $\Delta\Theta_{\textnormal{neighbours}}$ is fixed at \SI{0.05}{rad}. The search window size for nearest neighbours in radius $\Delta R_{\textnormal{neighbours}}$ is set to 75\% of the $l_{\textnormal{max}}$ of the current step.

\begin{table}
  \centering
  \caption{Overview of the configuration for the different steps of the pattern recognition chain. The last column shows some of the parameters of relevance for the CA as used for CLICdet:
  the maximum angle between cells $\alpha_{\textnormal{max}}$, the maximum cell length $l_{\textnormal{max}}$, 
  the minimum number of hits on track $N^{\textnormal{hits}}_{\textnormal{min}}$, the maximum $\chi^{2}_{\textnormal{max}}$ for valid track candidates, and $p_{\textnormal{T, cut}}$ used to discriminate between the two variations of the algorithm of track extension.}
  \label{tab:iterations}
  \begin{tabular}{cllccccc}\toprule
  \multirow{3}{*}{\textbf{Step}} & \multirow{3}{*}{\textbf{Algorithm}} & \multirow{3}{*}{\textbf{Hit collection}} & \multicolumn{5}{c}{\textbf{Parameters}}\\
  \cmidrule{4-8}
  & & & $\alpha_{\textnormal{max}}$ & $l_{\textnormal{max}}$ & $N^{\textnormal{hits}}_{\textnormal{min}}$ & $\chi^{2}_{\textnormal{max}}$ & $p_{\textnormal{T, cut}}$\\
  & & & [\SI{}{rad}] & [\SI{}{\mm}$^{-1}$] & - & -  & [\SI{}{\GeV}]\\
  \midrule
  0 & Building & Vertex Barrel & 0.005 & 0.02 & 4 & 100 & - \\ 
  1 & Extension & Vertex Endcap & 0.005 & 0.02 & 4 & 100 & 10 \\ 
  2 & Building & Vertex & 0.025 & 0.02 & 4 & 100 & - \\ 
  3 & Building & Vertex & 0.05 & 0.02 & 4 & 2000 & - \\ 
  4 & Extension & Tracker & 0.05 & 0.02 & 4 & 2000 & 1 \\ 
  5 \footnotesize{(default)} & Building & Vertex \& Tracker & 0.05 & 0.015 & 5 & 1000 & - \\ 
  5 \footnotesize{(isolated)} & Building & Vertex \& Tracker & 0.1 & 0.015 & 5 & 1000 & - \\
  \bottomrule
  \end{tabular}
\end{table}

Step 0 consists of the building of cellular track candidates using the hits in the vertex barrel as the seeding collection.
Step 1 performs the extension of cellular track candidates, found in step 0, limited to the set of hits in the vertex endcap.
 A further step of building of cellular track candidates is performed to recover remaining hits in the combined vertex barrel and endcap detectors, with tighter cuts at first to reconstruct higher-\pT tracks (step 2) and looser cuts afterwards to reconstruct low-\pT tracks (step 3).
All the candidates obtained via steps 0--3 are extended using hits in the tracker barrel and endcap detectors\footnote{The layer ordering of the tracker subdetectors, used in the track extension, is the following: inner tracker barrel (layers ordered by increasing radius), inner tracker endcap (layers ordered by increasing $z$ position), outer tracker barrel (layers ordered by increasing radius), outer tracker endcap (layer ordered by increasing $z$ position).} (step 4).   

\begin{figure}[htb]
\centering
\includegraphics[width=0.5\linewidth]{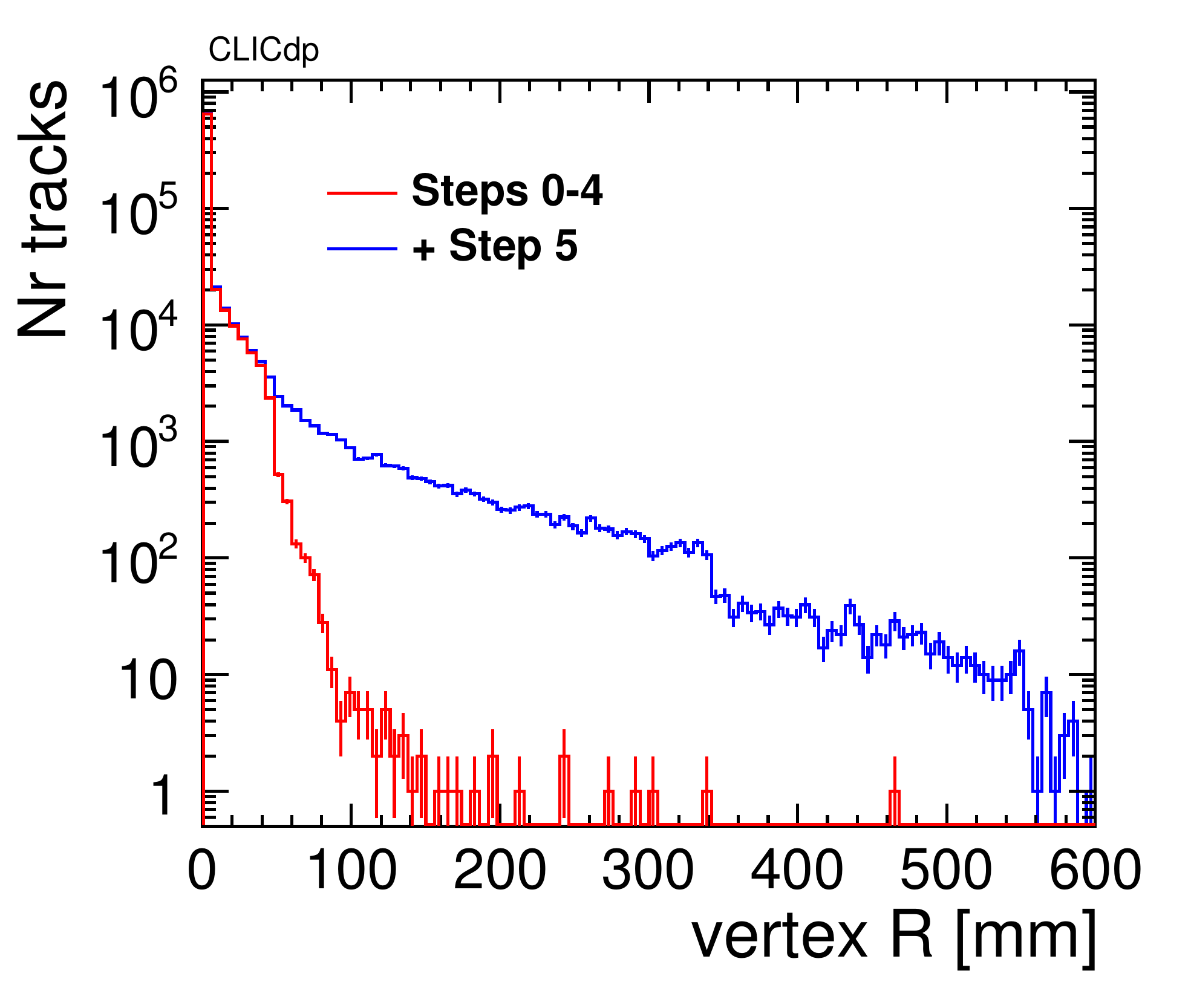}
\caption{Cumulative distribution of the total number of tracks reconstructed 
by the conformal tracking algorithm in CLICdet 
as a function of particle production vertex radius for step 0--4 (red) and including step 5 (blue).
The particles were produced in \epem$\rightarrow$ \ttbar events at \SI{3}{\TeV} centre-of-mass energy.}
\label{fig:ntracks_distr}
\end{figure}

Once steps 0--4 are completed, the unused hits in the entire tracking system can be used to reconstruct the remaining tracks,
most of which are generated by displaced particles as defined in~\ref{sec:CT_hitmapping}.
As shown in \cref{fig:conformal_mapping}, their trajectories are not mapped into straight lines in conformal space.
For this reason, the CA is executed with a special configuration:
\begin{itemize}
\item seed cells are formed at smaller conformal radius, i.e. in the outermost part of the tracker, and extended outwards in $(u,v)$ space;
\item the minimum number of hits required to form a track is increased with respect to the previous steps;
\item looser cuts in the CA parameters are set;
\item a quadratic term is included in the regression formula used to fit the cellular track candidates in conformal space, as was the case for the extension of low-\pT tracks.
\end{itemize}
The effect of this last step in the overall track reconstruction is clearly visible in~\cref{fig:ntracks_distr}
where the cumulative distribution of the total number of tracks reconstructed using different steps
in \epem$\rightarrow$ \ttbar events at \SI{3}{\TeV} centre-of-mass energy
is shown as a function of particle production vertex radius.
Given its special configuration, step 5 is the most computationally demanding part of the full pattern recognition chain, as described in \cref{subsec:cpu}.
Two configurations are summarised in \cref{tab:iterations}, as even looser cuts can be used when reconstructing single isolated particles without any drawback in terms of computing time. 
The comparison in terms of tracking efficiency between the two sets of cuts is discussed in~\cref{subsec:eff_fake}.

\begin{figure}[htb]
\centering

  \begin{subfigure}{0.45\textwidth}
  \includegraphics[width=\linewidth]{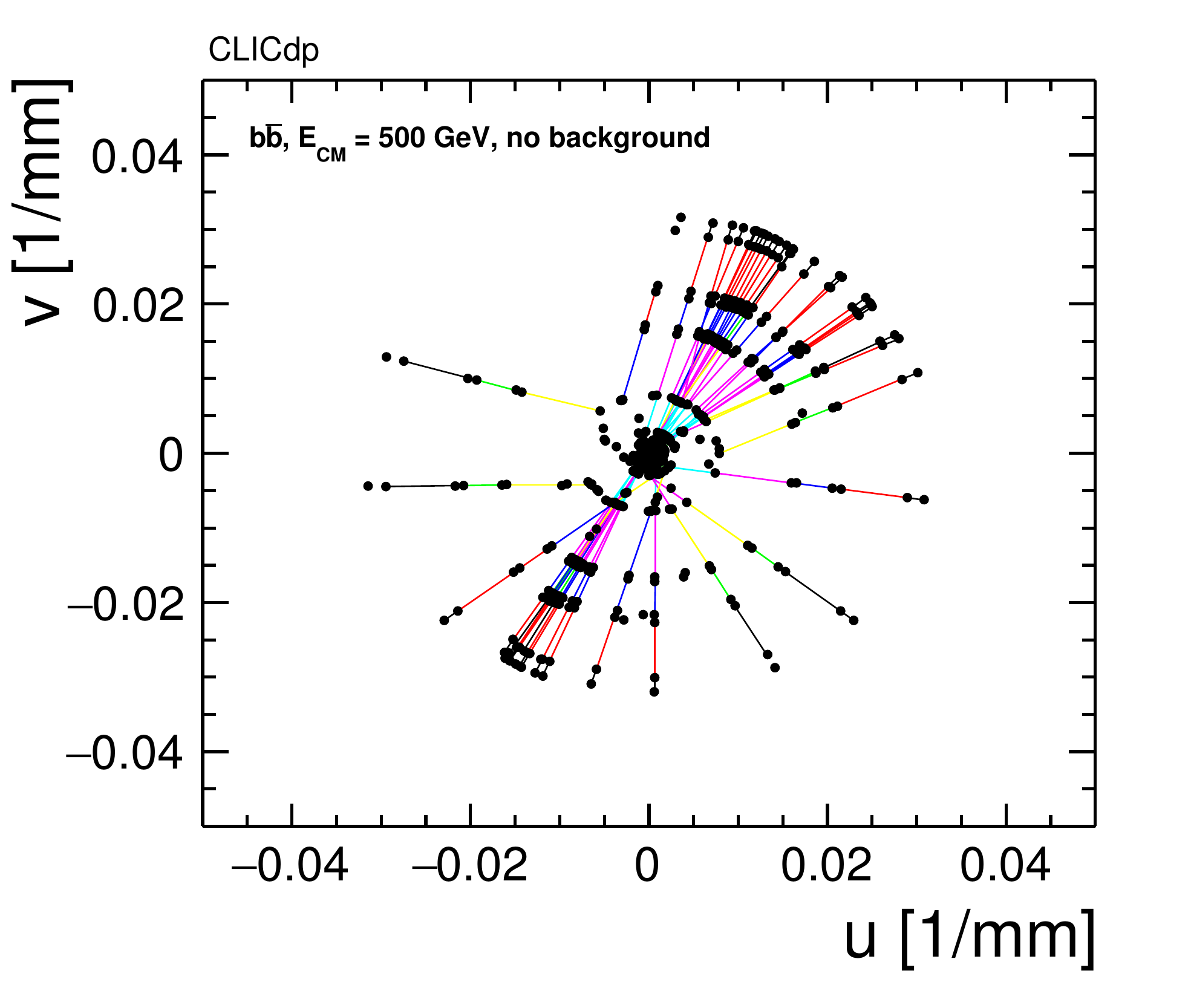}
  \end{subfigure}\hfil
  \begin{subfigure}{0.45\textwidth}
  \includegraphics[width=\linewidth]{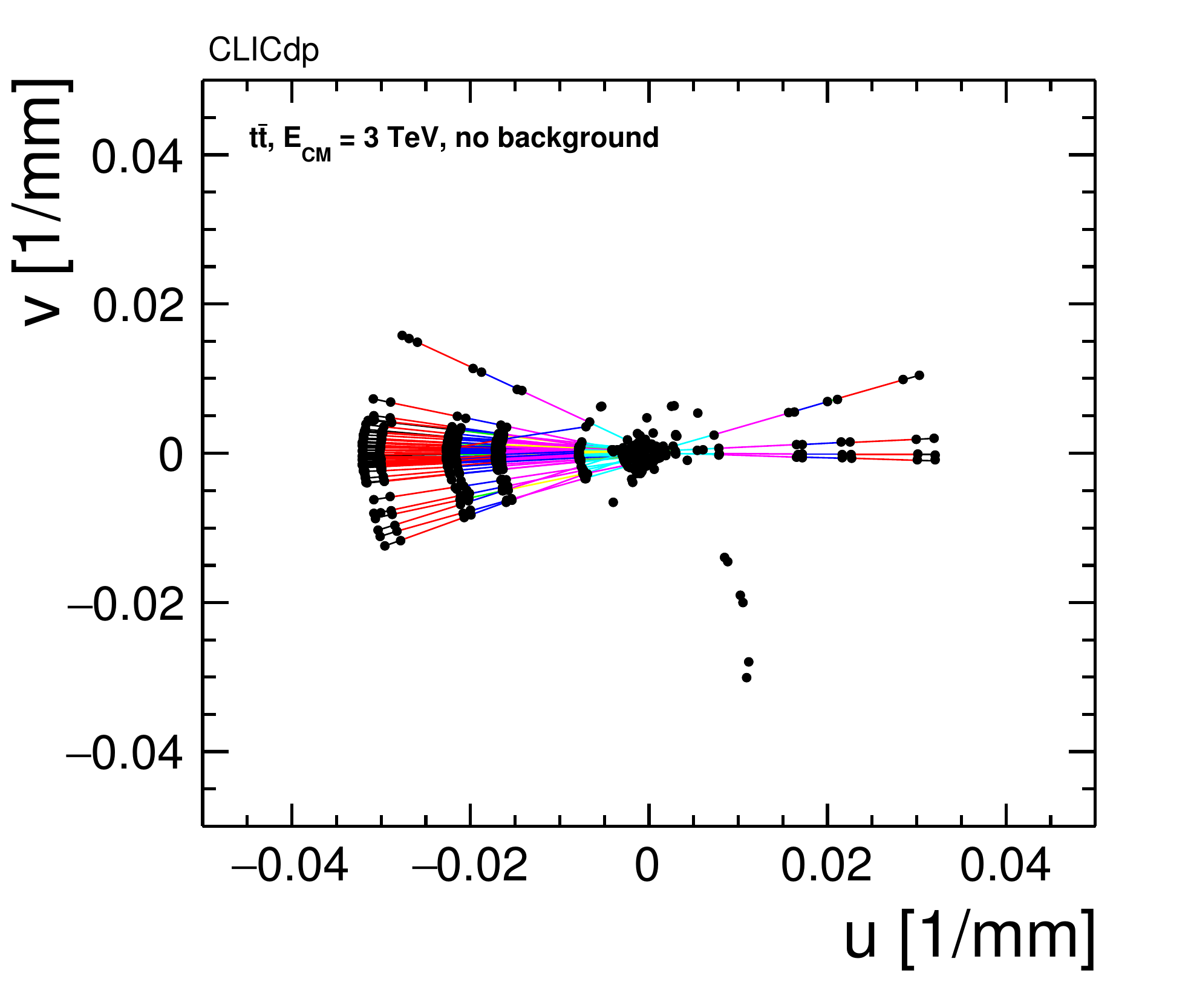}
  \end{subfigure}\hfil

  \caption{Cellular track candidates reconstructed by the conformal tracking algorithm in CLICdet in the case of a single \bb event at \SI{500}{\GeV} centre-of-mass energy (left) and
  a single \ttbar event at \SI{3}{\TeV} (right).}
  \label{fig:CAtracks}
\end{figure}

In~\cref{fig:CAtracks} all cellular track candidates reconstructed by the full pattern recognition chain are shown
for two example events: a single \bb event at \SI{500}{\GeV} centre-of-mass energy and a single \ttbar event at \SI{3}{\TeV}.
The performance in terms of track finding in a large \epem$\rightarrow$ \ttbar sample at \SI{3}{\TeV} 
centre-of-mass energy is studied in detail in~\cref{sec:performance}.

\subsubsection{Track fitting}
\label{subsec:track_fitting}

For each cellular track candidate reconstructed during the pattern recognition stage,
the track fit is run with the aim of obtaining a precise estimation of the track parameters.
It consists of a Kalman filter and smoother in the global $(x,y,z)$ coordinate space~\cite{Fruehwirth:KF}.

As a first step, a simple helix is fitted to three hits of the track (typically first, middle and last hit) to obtain
a first estimation of the trajectory parameters. This step is also called \emph{pre-fit}.
Different combinations of hits along a track were tested to prove the robustness of this method.

The parameters obtained from the pre-fit are used to initialise the Kalman filter,
while the values of the corresponding covariance matrix are set several order of magnitudes larger than the expected ones
to perform an unbiased fit.
The Kalman filter and smoother is then run as implemented in the KalTest package described in~\cite{ILCsoft:KF}.
It proceeds through the full list of hits on the track from the innermost to the outermost layer of the CLICdet tracking system, 
updating the track trajectory state, which contains the estimation of track parameters and their uncertainties, sequentially with each hit.
In the smoother step, a second filter is initialised with the result of the first one
and is run from the outermost layer to the innermost.
In case the fit fails, the Kalman filter and smoother is attempted a second time using the hits 
collected in the tracks in reverse order.

Some of the hits on the cellular track candidates may be rejected during the track fitting step, 
based on requirements on the $\chi^{2}_{tot}$ set in the track fit itself. 
As a result, some tracks can have fewer hits than the minimum requirement imposed during the pattern recognition stage. 
Dedicated studies in \bb and \ttbar events have shown that short tracks are in general of bad quality.
Therefore they are filtered out in a later stage, as described in~\cref{subsec:track_selection}.

\subsubsection{Track selection}
\label{subsec:track_selection}

The track finding procedure can yield a significant fraction of undesirable tracks, such as
tracks that contain many spurious hits or share a number of hits with other tracks. 
Therefore, a track selection is applied as the final stage with the goal of reducing the number of these tracks while keeping high tracking performance.

In order to avoid tracks with more than two shared hits, the clone treatment described in~\cref{sec:CA_build} 
based on the track length and $\chi^{2}$ is repeated on the fitted track collection. 
In this case, the $\chi^{2}$ used to determine the highest quality track is the one calculated 
at the end of the Kalman filter and smoother procedure. 

As mentioned in \cref{subsec:track_fitting}, introducing a lower threshold on the total number of hits belonging to a track
is an effective way to reduce the number of bad-quality tracks. 
The optimal minimum number of hits has been found to be equal to the minimum imposed on the first building step, i.e. 4.
However, this is not entirely the case for the reconstruction of single isolated particles, for which rejecting three-hit tracks 
can result in efficiency loss with limited improvement on the track quality.
Therefore, the requirement of number of hits larger than three is applied to the tracks as an offline requirement according to the event topology.

\subsection{Performance}
\label{sec:performance}

In this section, the performance of the track reconstruction process used in the CLIC environment 
is investigated in terms of tracking efficiency and mis-reconstruction, 
track parameter resolutions, and CPU time required for the track reconstruction.
To assess the track mis-reconstruction, the fraction of spurious tracks reconstructed in each event, also known as fake rate, is computed.
The efficiency and fake rate are evaluated for simulated isolated particles and realistic topologies in a high occupancy environment.
Track parameter resolutions are extracted using simulated single muons. 

Results for tracking efficiency and fake rate are presented in \cref{subsec:eff_fake}. Resolutions of track parameters are examined in \cref{subsec:res}. Finally, \cref{subsec:cpu} provides an analysis of the CPU time required for the different steps of the tracking algorithm.

\subsubsection{Tracking efficiency and fake rate} 
\label{subsec:eff_fake}

Reconstructed tracks are matched with simulated particles to evaluate the tracking efficiency and fake rate. 
The simulated particle from which the majority of track hits originate is defined as the \emph{associated} particle.
Track \emph{purity} is defined as the fraction of track hits which origin from the associated particle.

\begin{figure}[htb]
\centering
\includegraphics[width=0.5\linewidth]{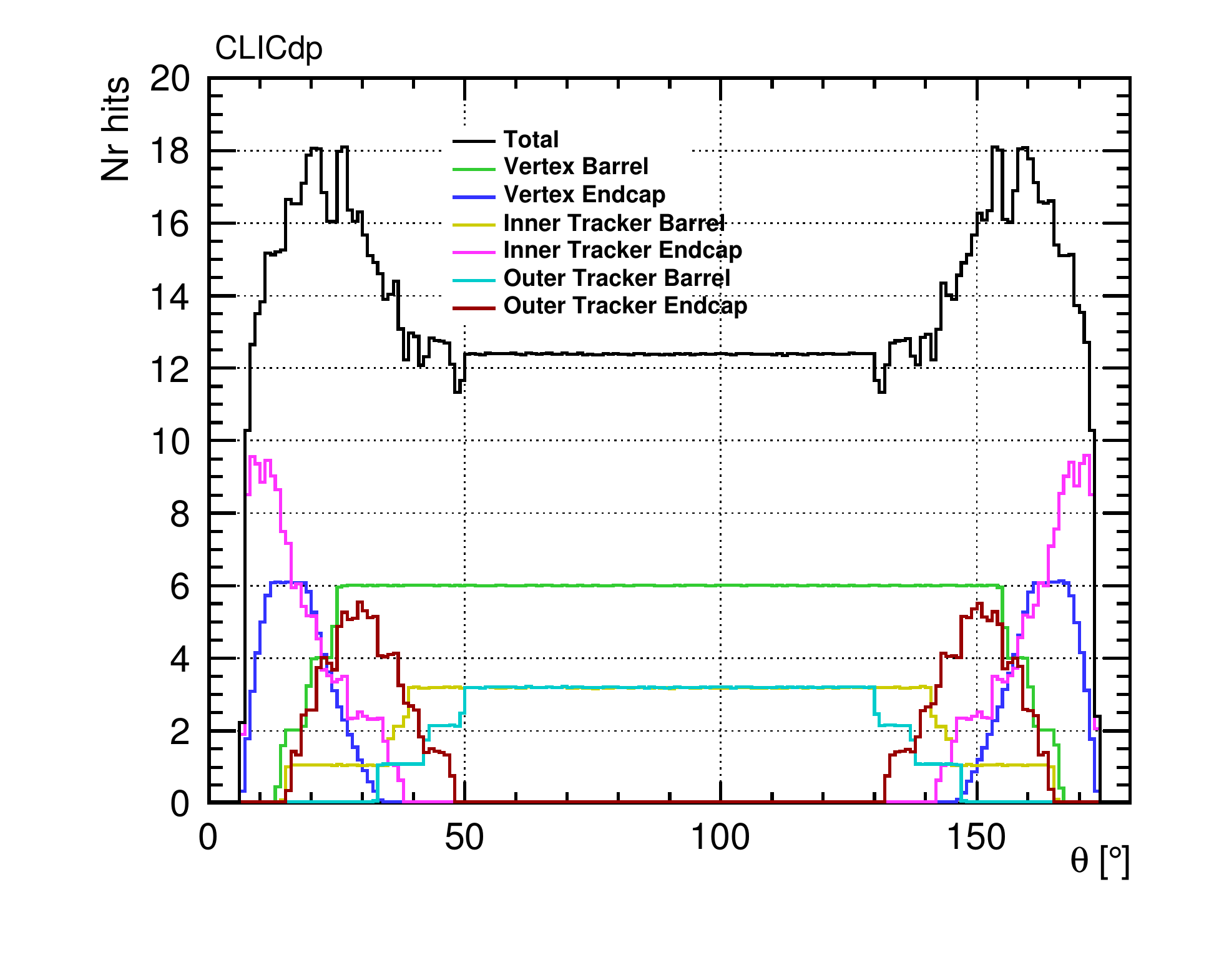}
\caption{Distribution of the number of hits produced by a simulated particle crossing the entire tracking system in CLICdet.}
\label{fig:nhits_distr}
\end{figure}

A simulated particle is considered reconstructable if the following requirements are satisfied: 
\pT > \SI{100}{\MeV}, $|\cos(\theta)|$ < 0.99, at least 4 hits on different layers
and does not decay in the tracker volume.
The requirement on the polar angle guarantees that prompt tracks with negligible energy loss cross at least ten sensitive layers in the CLICdet tracker,
as shown in~\cref{fig:nhits_distr}.

The tracking \emph{efficiency} is defined as the ratio of reconstructed tracks which can be associated to at least one 
reconstructable particle with at least 75\% purity over the total number of reconstructable particles.
In case the reconstructed track has a purity of less than 75\%, it is classified as fake.
The \emph{fake rate} is therefore defined as the fraction of reconstructed tracks that are classified as fake.

The efficiency is presented as a function of the \pT and polar angle $\theta$ of the associated simulated particles,
while in the case of the fake rate, the \pT and polar angle $\theta$ of the reconstructed track is used.

\paragraph{Results for isolated particles}

\begin{figure}[htb]
\centering

\begin{subfigure}{0.45\textwidth}
\includegraphics[width=\linewidth]{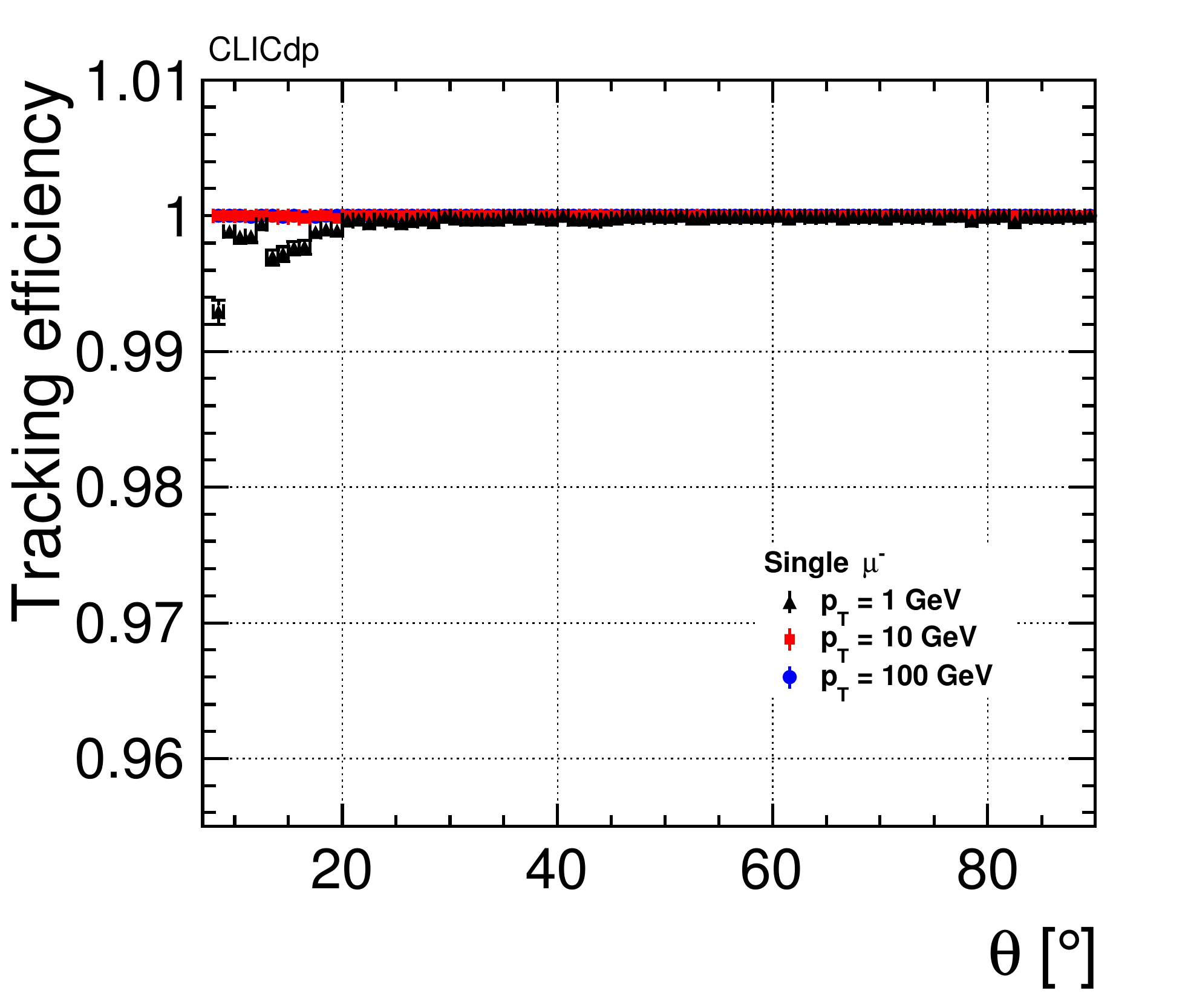}
\end{subfigure}\hfil
\begin{subfigure}{0.45\textwidth}
\includegraphics[width=\linewidth]{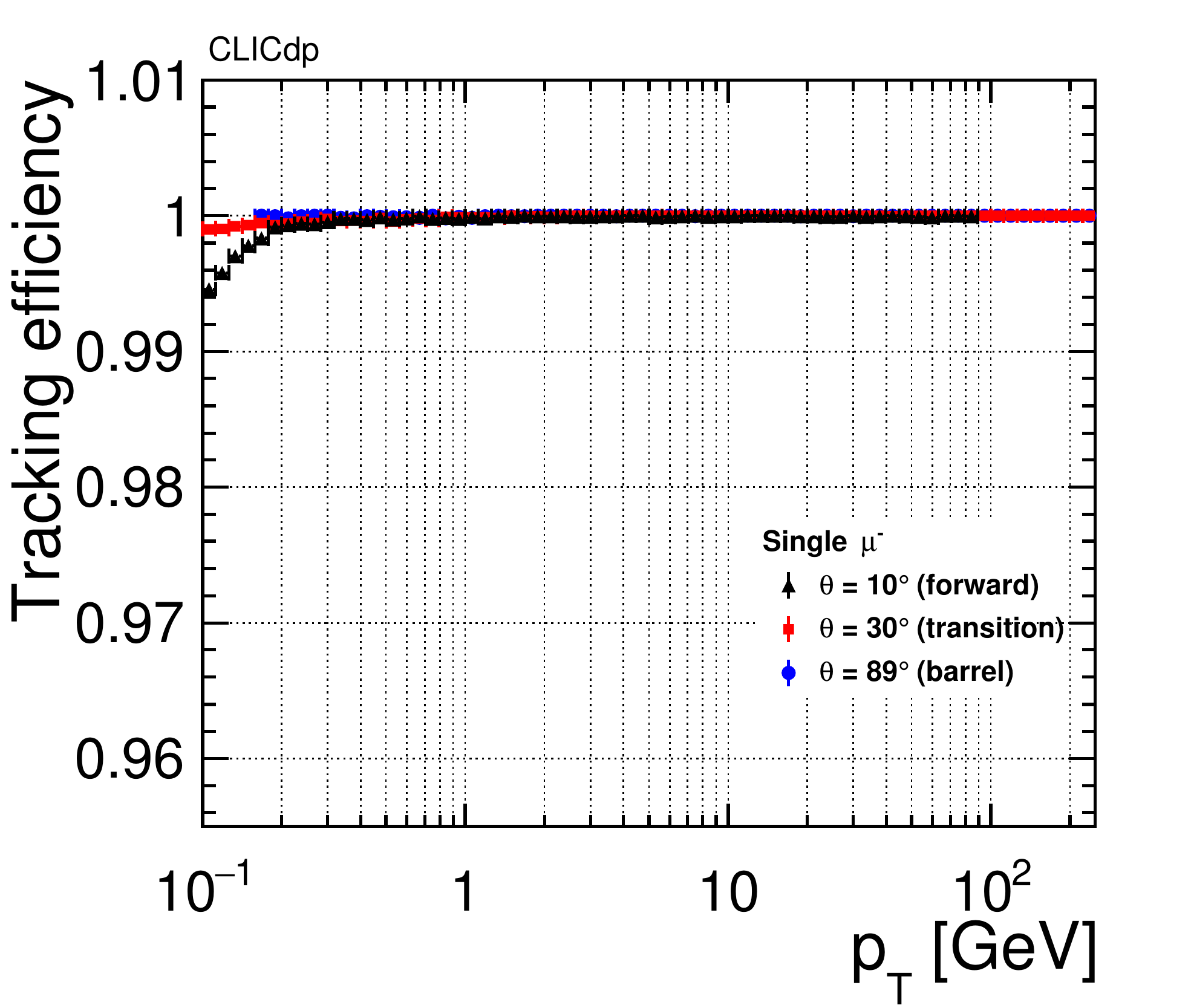}
\end{subfigure}\hfil

\medskip
\begin{subfigure}{0.45\textwidth}
\includegraphics[width=\linewidth]{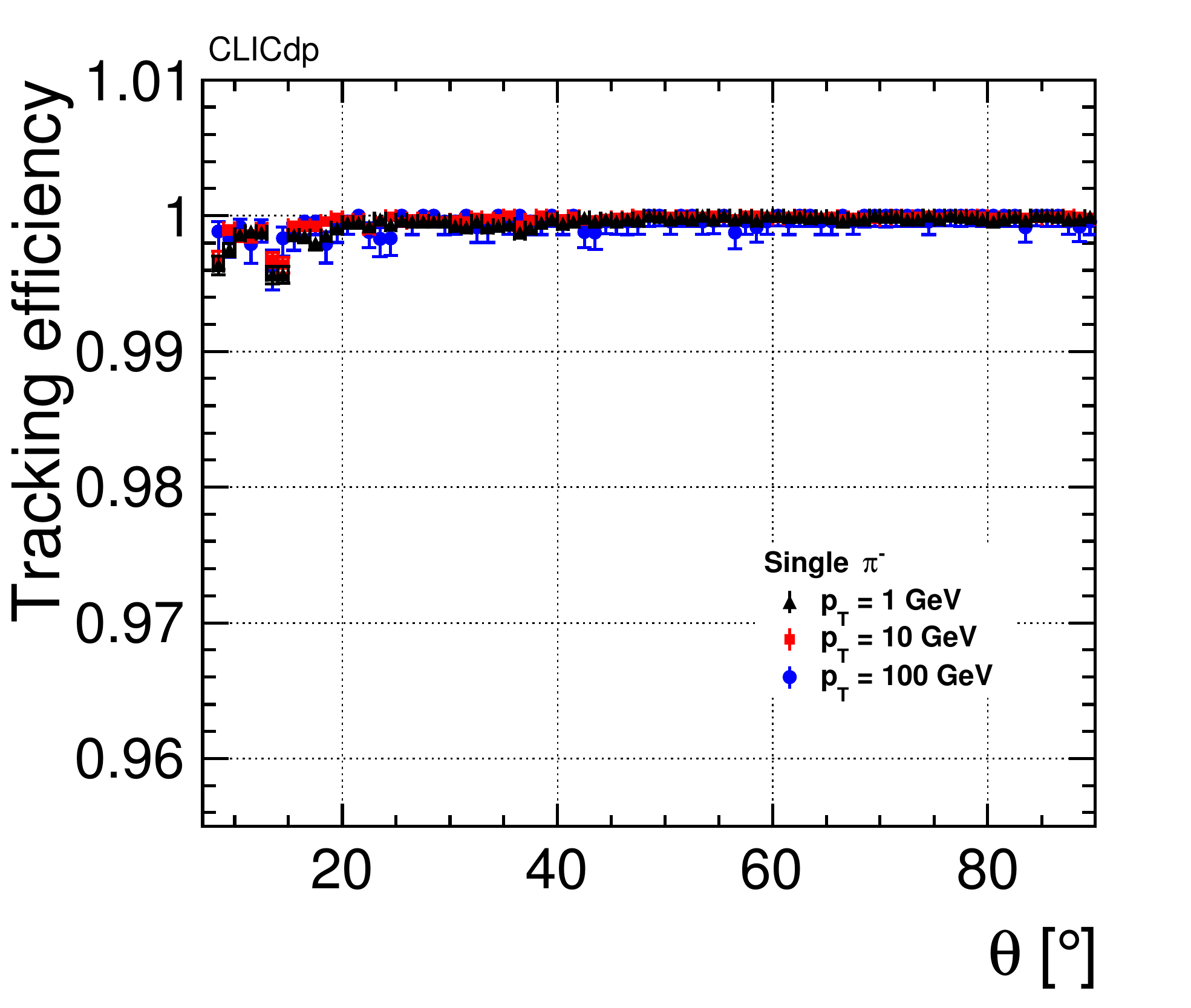}
\end{subfigure}\hfil
\begin{subfigure}{0.45\textwidth}
\includegraphics[width=\linewidth]{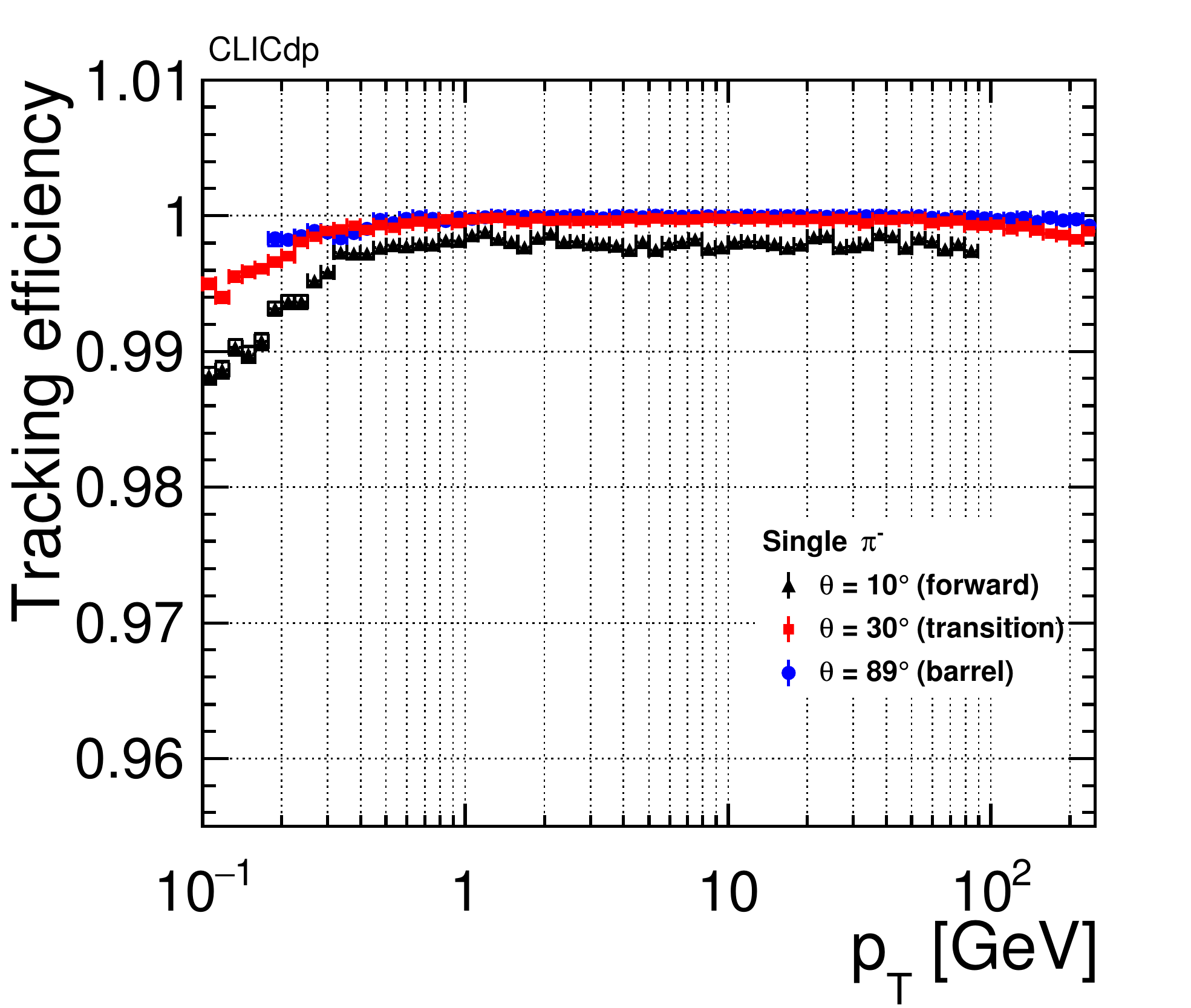}
\end{subfigure}\hfil

\medskip
\begin{subfigure}{0.45\textwidth}
\includegraphics[width=\linewidth]{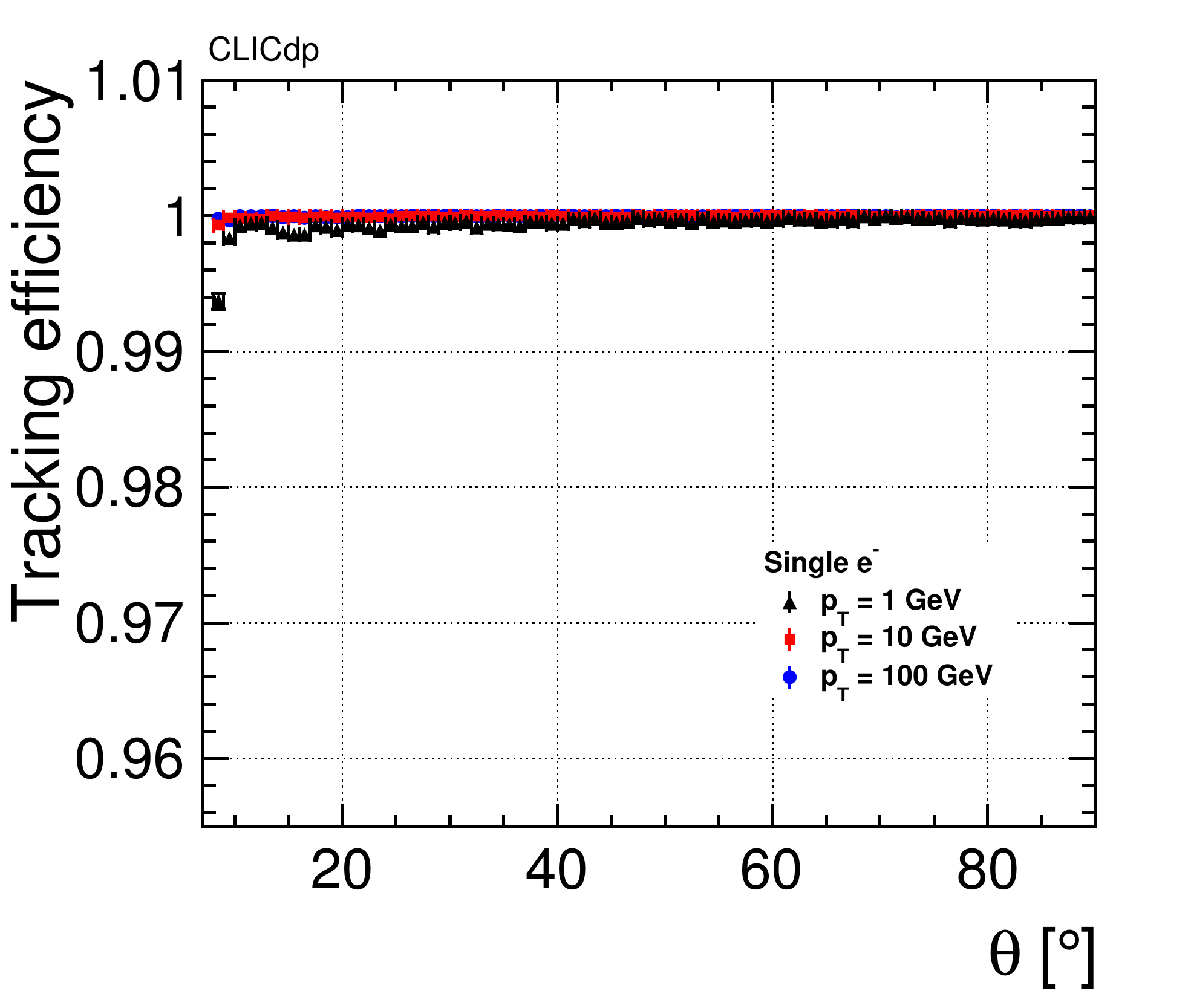}
\end{subfigure}\hfil
\begin{subfigure}{0.45\textwidth}
\includegraphics[width=\linewidth]{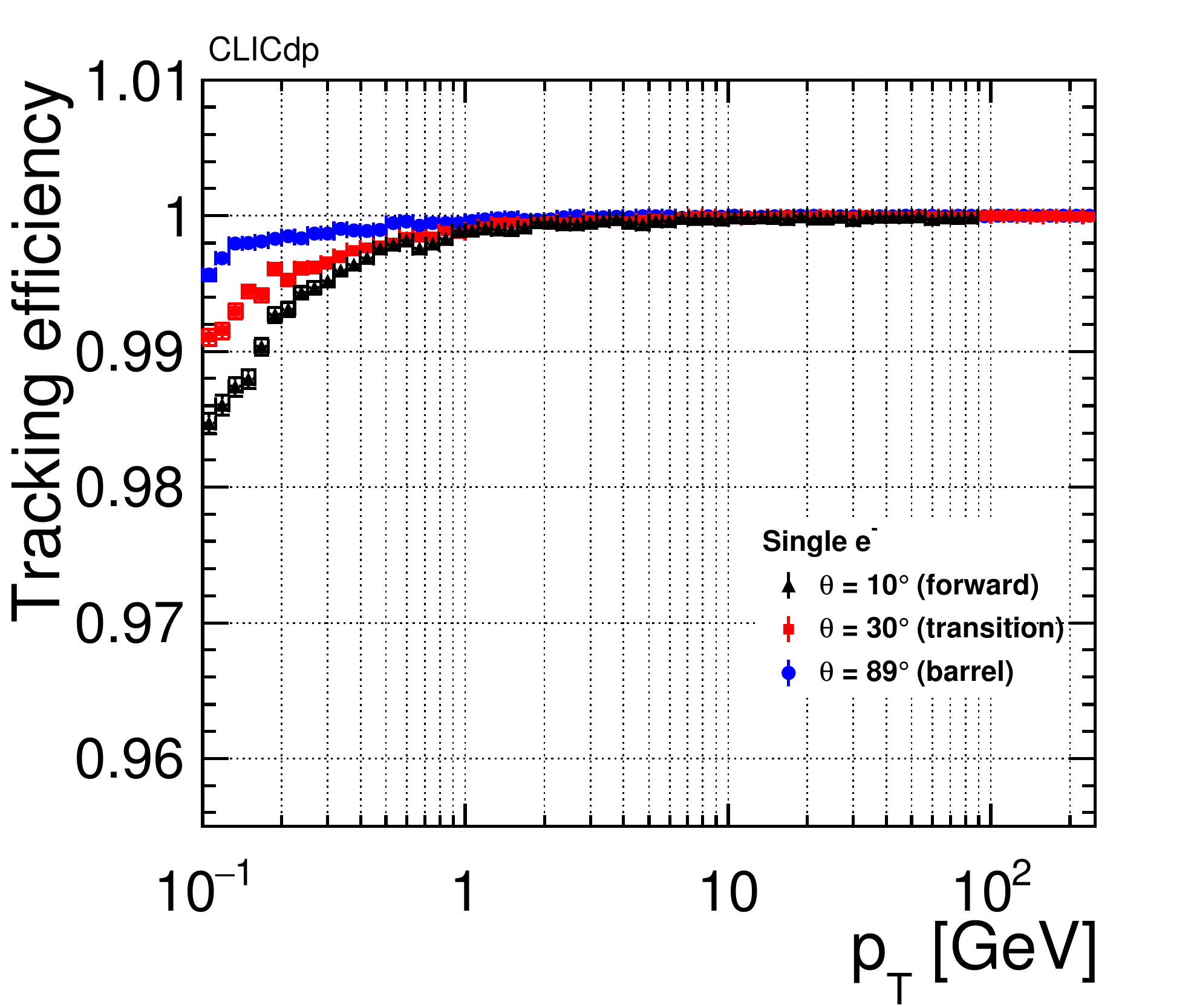}
\end{subfigure}

\caption{Track reconstruction efficiency for isolated muons (top), pions (middle), and electrons (bottom). Results are shown as a function of polar angle $\theta$ for \pT = \SI{1}{\GeV}, \SI{10}{\GeV}, and \SI{100}{\GeV} (left) and as a function of \pT for $\theta$ = \ang{10}, \ang{30}, and \ang{89}(right).}
\label{fig:singlePart_eff}
\end{figure}

Isolated muons are the easiest topology to reconstruct as they interact with the detector material almost exclusively through ionization, 
their energy loss via bremsstrahlung being generally negligible. Therefore, they cross the entire detector without large deviations from a helical trajectory.
Moreover, both the multiple Coulomb scattering and energy loss through ionization are properly implemented in the Kalman filter.
For isolated muons, the tracking is fully efficient in the entire tracker acceptance and for the considered transverse momentum (\cref{fig:singlePart_eff}, top). 
The fake rate is negligible.

\begin{figure}[htb]
\centering

\begin{subfigure}{0.45\textwidth}
\includegraphics[width=\linewidth]{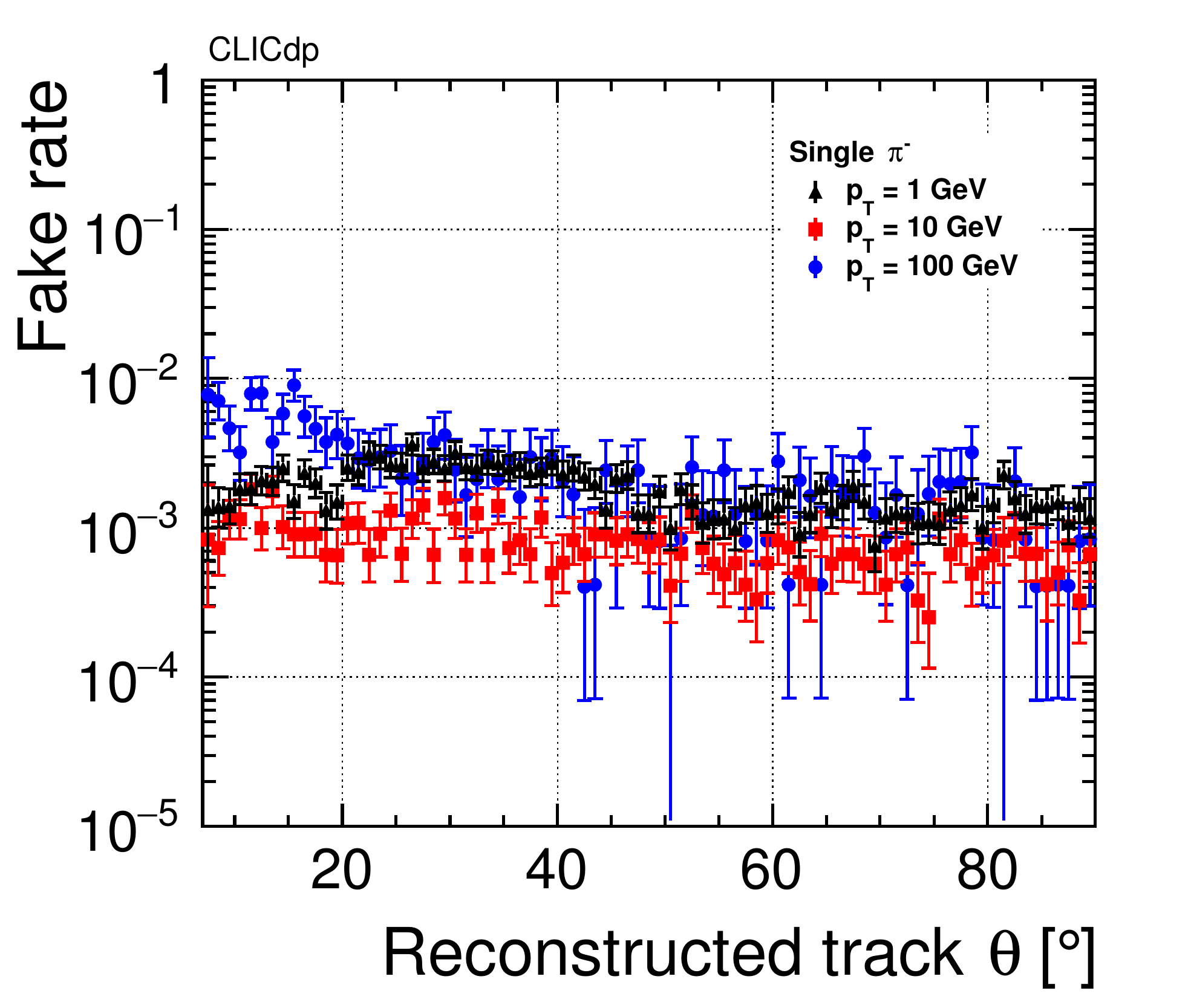}
\end{subfigure}\hfil
\begin{subfigure}{0.45\textwidth}
\includegraphics[width=\linewidth]{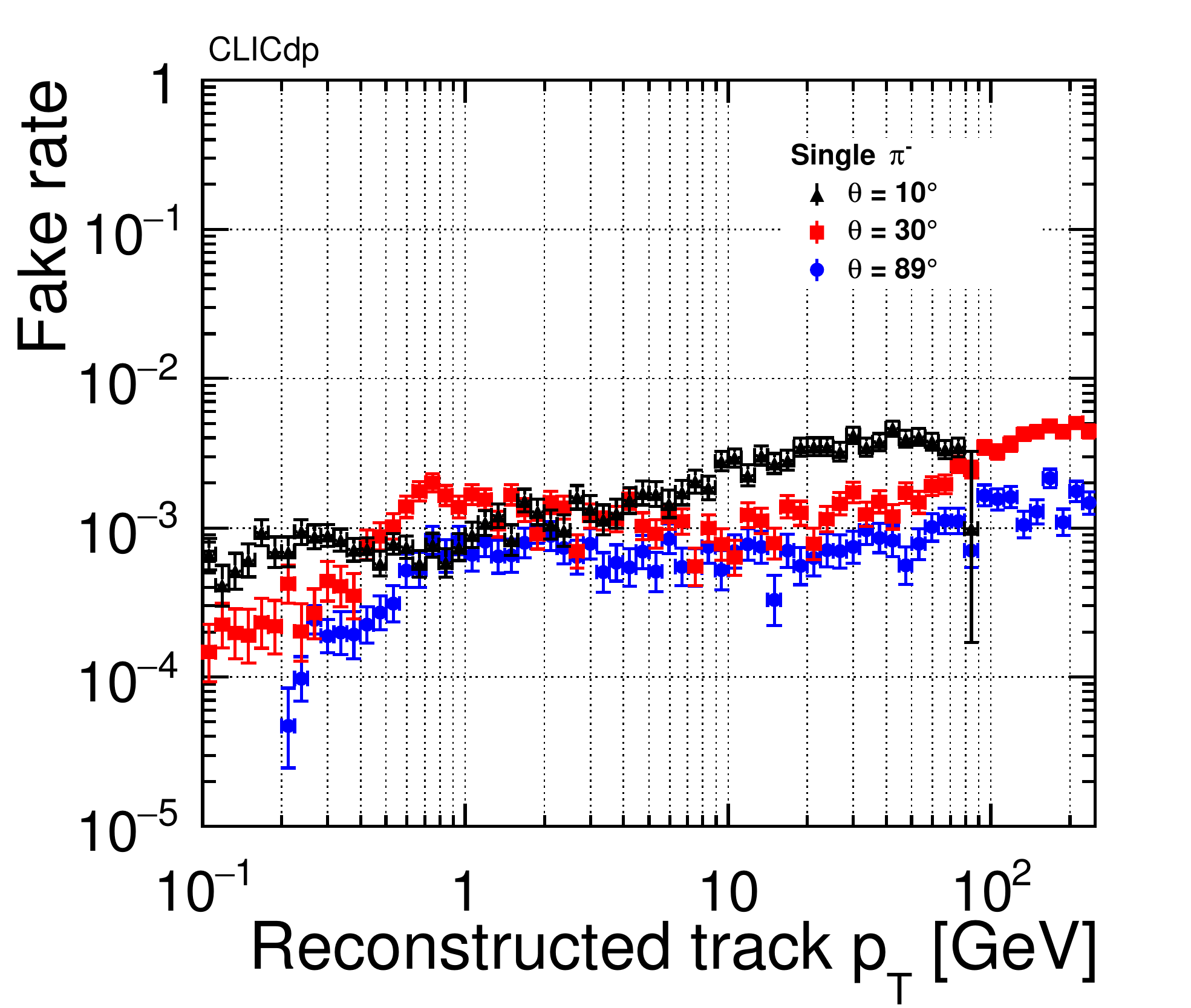}
\end{subfigure}\hfil

\medskip
\begin{subfigure}{0.45\textwidth}
\includegraphics[width=\linewidth]{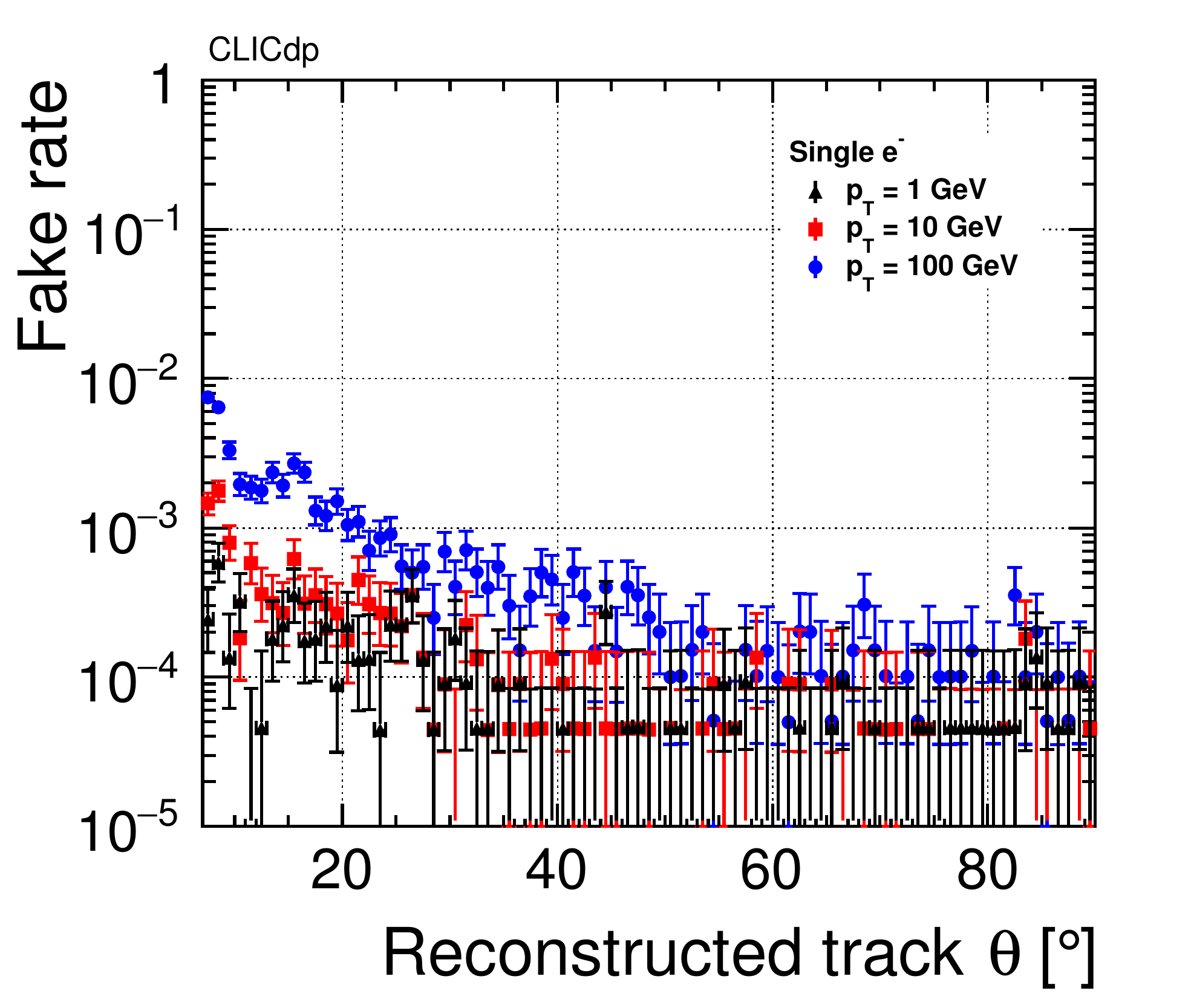}
\end{subfigure}\hfil
\begin{subfigure}{0.45\textwidth}
\includegraphics[width=\linewidth]{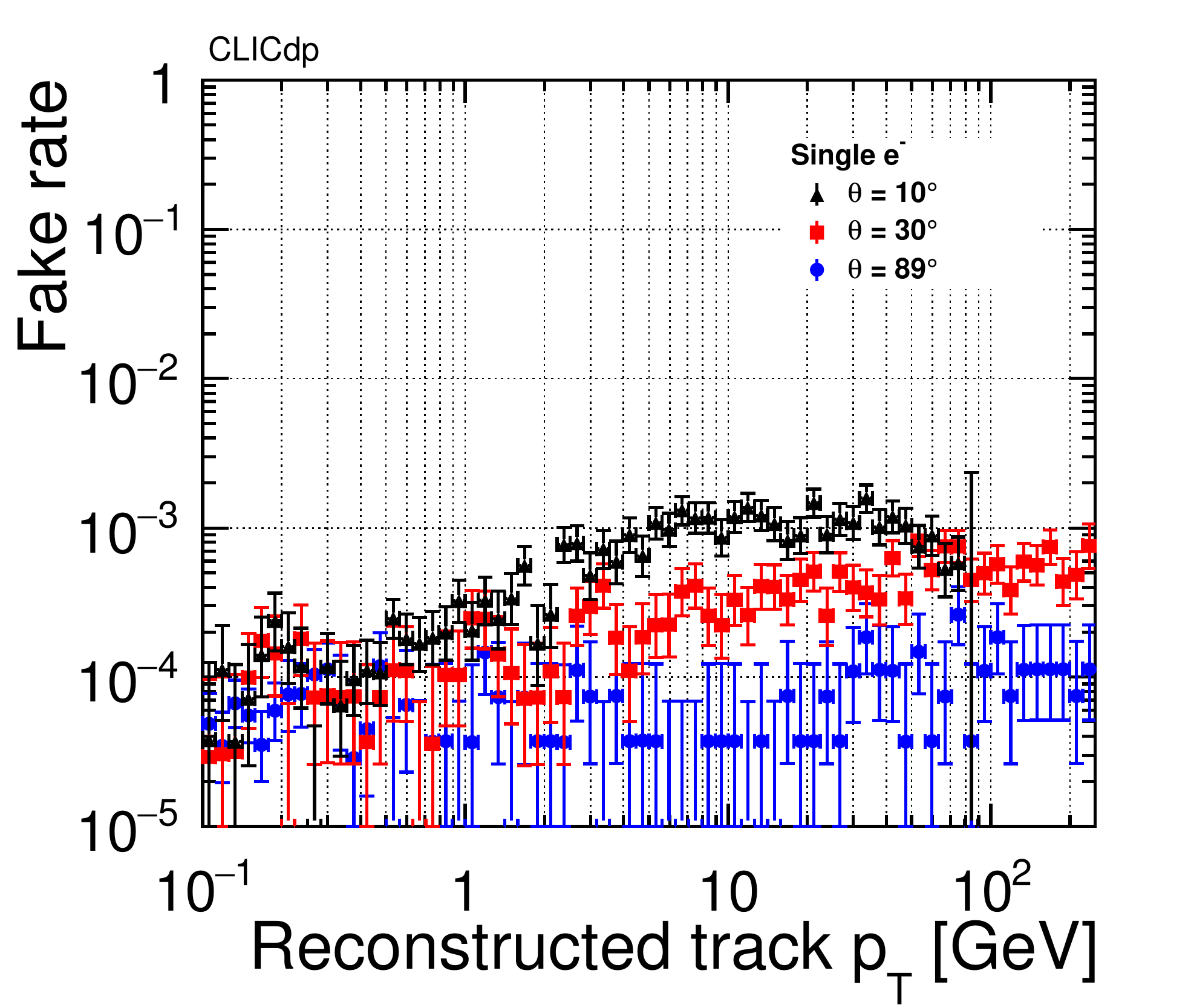}
\end{subfigure}

\caption{Tracking fake rate for isolated pions (top) and electrons (bottom). Results are shown as a function of polar angle $\theta$ for \pT = \SI{1}{\GeV}, \SI{10}{\GeV}, and \SI{100}{\GeV} (left) and as a function of \pT for $\theta$ = \ang{10}, \ang{30}, and \ang{89}(right). NB: the measured fake rate depends strongly on the \pT distribution of the simulated particles: if no particles are produced in a given \pT range, most tracks reconstructed in that range must be fake. Therefore, only results within the \pT range of simulated particles are shown.}
\label{fig:singlePart_fake}
\end{figure}

The reconstruction of charged pions is more complicated, because of hadronic interactions. In particular, the deviation from the Coulomb scattering angle caused by elastic nuclear interactions is not taken into account in the Kalman filter. This may result in reconstructed tracks with fewer hits than the simulated particle or in so-called \emph{split tracks}, i.e. two reconstructed parts of one particle trajectory.
For isolated charged pions, the tracking is fully efficient for particles with \pT down to approximately \SI{600}{\MeV} in the barrel and transition region, while an efficiency loss at the permille level is observed in the forward region (\cref{fig:singlePart_eff}, middle). 
The small dip observed at \ang{15} is caused by the slightly larger material budget traversed by particles in this direction.
The fake rate is well below 1\% for all transverse momenta and slightly increases with \pT due to the fact that in an inelastic hadron interaction primary and secondary particles are
more collimated and therefore hits coming from secondary particles are more likely to be included in the primary track as spurious hits (\cref{fig:singlePart_fake}, top).

Electrons in the investigated momentum range lose energy mainly through bremsstrahlung. This may cause the electrons not to reach the outer layers, thus reducing the number of particle hits. Moreover, radiated photons can convert into \epem pairs, whose hits could be attached to the primary track. 
The tracking efficiency for isolated electrons is 100\% for transverse momenta above \SI{1}{\GeV}. A maximum efficiency loss of 1--2\% is observed for lower-\pT particles, depending on their polar angle (\cref{fig:singlePart_eff}, bottom). 
The fake rate is of the order of permille for particles in the transition and forward region, where the tracker material budget is larger, and one order of magnitude smaller for particles in the barrel (\cref{fig:singlePart_fake}, bottom).

\begin{figure}[tb]
\centering

\begin{subfigure}{0.45\textwidth}
\includegraphics[width=\linewidth]{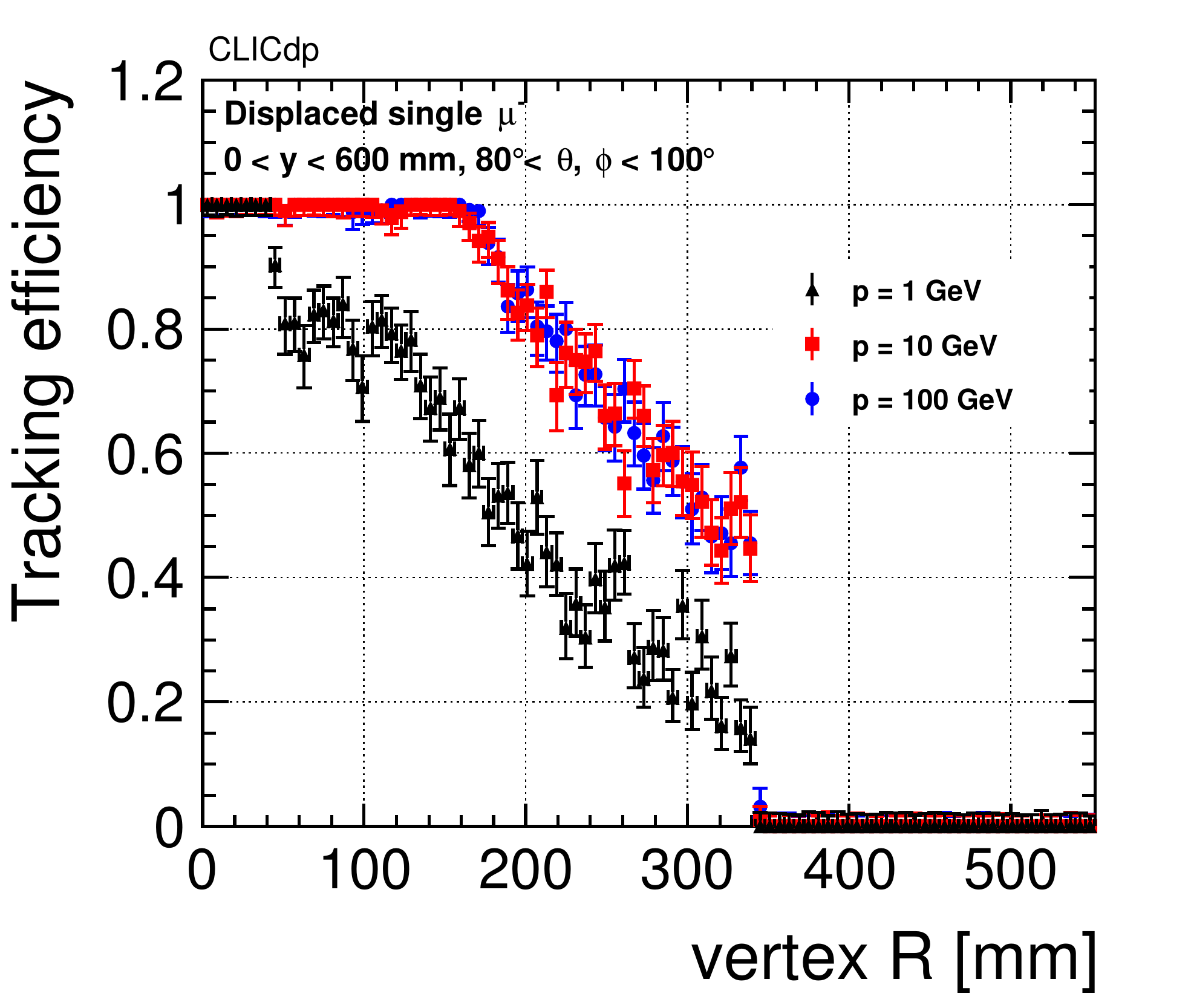}
\end{subfigure}\hfil
\begin{subfigure}{0.45\textwidth}
\includegraphics[width=\linewidth]{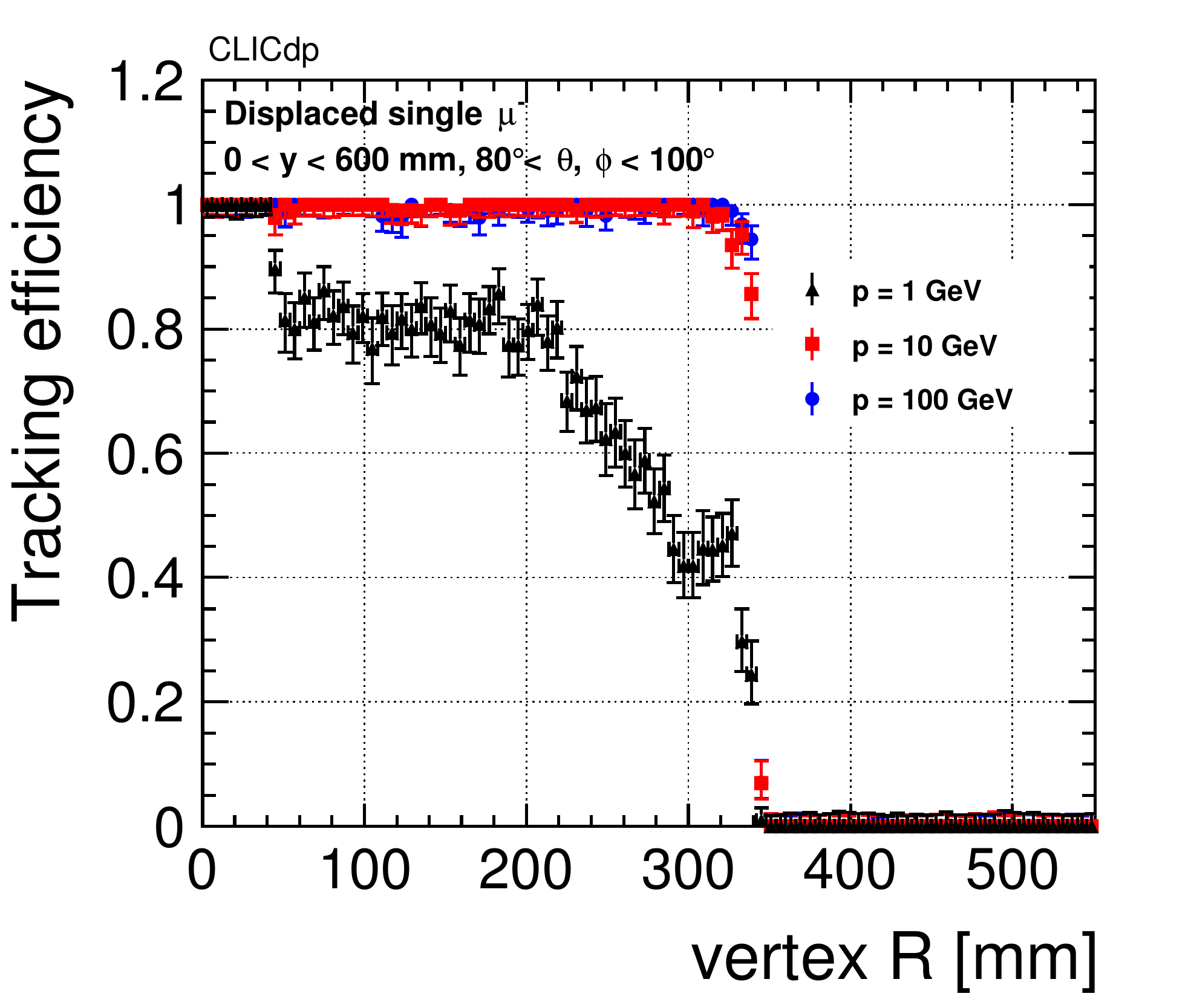}
\end{subfigure}\hfil

\caption{Tracking efficiency for isolated displaced muons with momenta of \SI{1}{\GeV}, \SI{10}{\GeV}, and \SI{100}{\GeV} as a function of production vertex radius with two sets of reconstruction requirements: default (left) and tuned to the isolated particle events (right). The sample is generated with a displacement of up to \SI{60}{\cm} in the $y$ coordinate and an uniform polar angle distribution in a \ang{10} cone around the $y$ axis.}
\label{fig:singlePartDisplaced}
\end{figure}

The sample used for studying displaced particles as defined in~\cref{sec:CT_hitmapping} consists of isolated muons generated 
with a displacement of up to \SI{60}{\cm} in the $y$ coordinate and with a polar angle uniformly distributed in a \ang{10} cone around the $y$ axis.
Muons, which do not undergo strong interactions, have been generated with this distribution in order to facilitate the interpretation of the results: the number of hits per particle corresponds to the number of barrel layers with radii greater than or equal to the particle production vertex radius\footnote{In the specific simulated sample, the particle production vertex radius $R$ is equal to the displacement in $y$.}.
The result is shown in \cref{fig:singlePartDisplaced} for muons with momenta of \SI{1}{\GeV}, \SI{10}{\GeV}, and \SI{100}{\GeV} as a function of the particle production vertex radius, comparing the effect of the two sets of requirements introduced in~\cref{subsubsec:chain}: default (left) and tuned to the isolated particle events (right). This shows the flexibility of the algorithm in adapting the thresholds according to the topology of the studied sample.
When using the tuned set of cuts, the tracking is fully efficient for the \SI{10}{\GeV} and \SI{100}{\GeV} particles, up to a production radius of \SI{35}{\cm}. Particles produced with a larger displacement do not cross the minimum number of layers required in the CA. For muons with momenta of \SI{1}{\GeV}, the efficiency is 100\% as long as the particles are produced before the second double layer of the vertex detector (radius smaller than \SI{4}{\cm}). 

\paragraph{Results for topologies in high occupancy environment}

Reconstructing tracks in jet topologies is more challenging than in events with single isolated particles. 
Simulated \epem$\rightarrow$ \ttbar events at \SI{3}{\TeV} centre-of-mass energy are used to probe the dependence of the tracking 
performance on particle proximity in very collimated jets. 
In addition, the overlay of \gghadron background expected at the \SI{3}{\TeV} energy stage increases further the hit occupancy, 
thus the complexity of the pattern recognition.

\begin{figure}[htb]
\centering

\begin{subfigure}{0.45\textwidth}
\includegraphics[width=\linewidth]{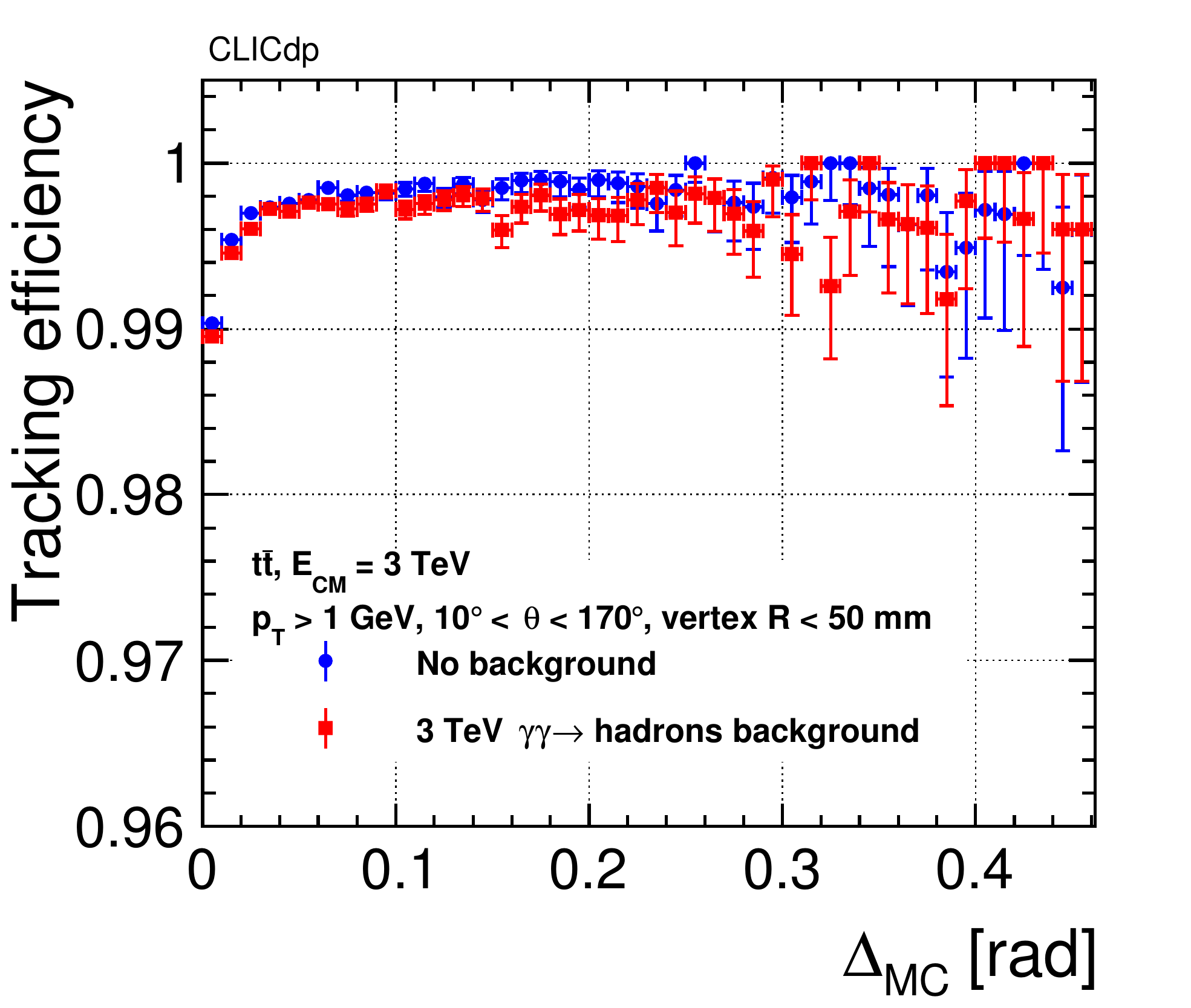}
\end{subfigure}\hfil
\begin{subfigure}{0.45\textwidth}
\includegraphics[width=\linewidth]{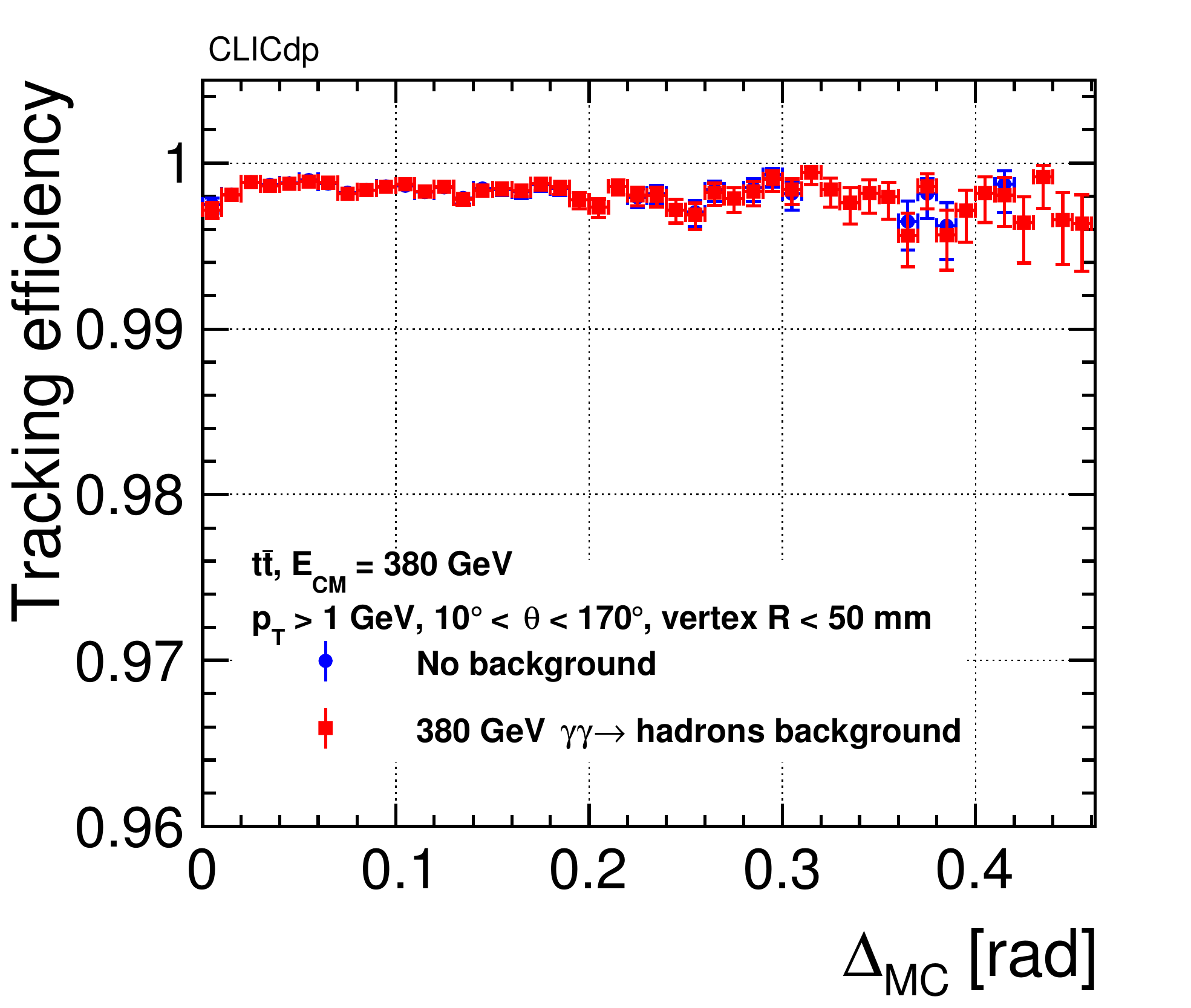}
\end{subfigure}\hfil

\caption{Tracking efficiency as a function of particle proximity for \epem$\rightarrow$ \ttbar events at \SI{3}{\TeV} (left) and \SI{380}{\GeV} (right) centre-of-mass energy. Results are shown for samples reconstructed with and without the overlay of 30 bunch crossings of
\gghad background expected at the \SI{3}{\TeV} and \SI{380}{\GeV} CLIC energy stage, respectively.}
\label{fig:ttbar_dmc}
\end{figure}

Results are presented in terms of tracking efficiency as a function of particle proximity $\Delta_{\textnormal{MC}}$ (\cref{fig:ttbar_dmc}), defined as the minimum distance between the particle associated with the track and any other particle\footnote{$\Delta_{\textnormal{MC}}$ is computed as $\sqrt{(\Delta\eta)^{2}+(\Delta\phi)^{2}}$, where $\eta$ is the pseudorapidity defined as $-\ln[\tan(\theta/2)]$.}. Pure tracks as defined in \cref{subsec:eff_fake} have been used with the following requirements: \ang{10} < $\theta$ < \ang{70}, \pT > \SI{1}{\GeV}, production radius smaller than \SI{50}{\mm}. The maximum efficiency loss, observed for $\Delta_{\textnormal{MC}}$ < \SI{0.02}{\rad}, amounts to 1\%. 
For comparison, performance for events with the same topology but \SI{380}{\GeV} centre-of-mass energy are shown (\cref{fig:ttbar_dmc}, right). 
Also in this case, an efficiency loss for $\Delta_{\textnormal{MC}}$ < \SI{0.02}{\rad} is present. 
In order to study the performance of the tracking algorithm in less extreme occupancy conditions than $\Delta_{\textnormal{MC}}$ < \SI{0.02}{\rad}, those particles are excluded from all following analyses. 

\begin{figure}[tbh]
\centering

\begin{subfigure}{0.45\textwidth}
\includegraphics[width=\linewidth]{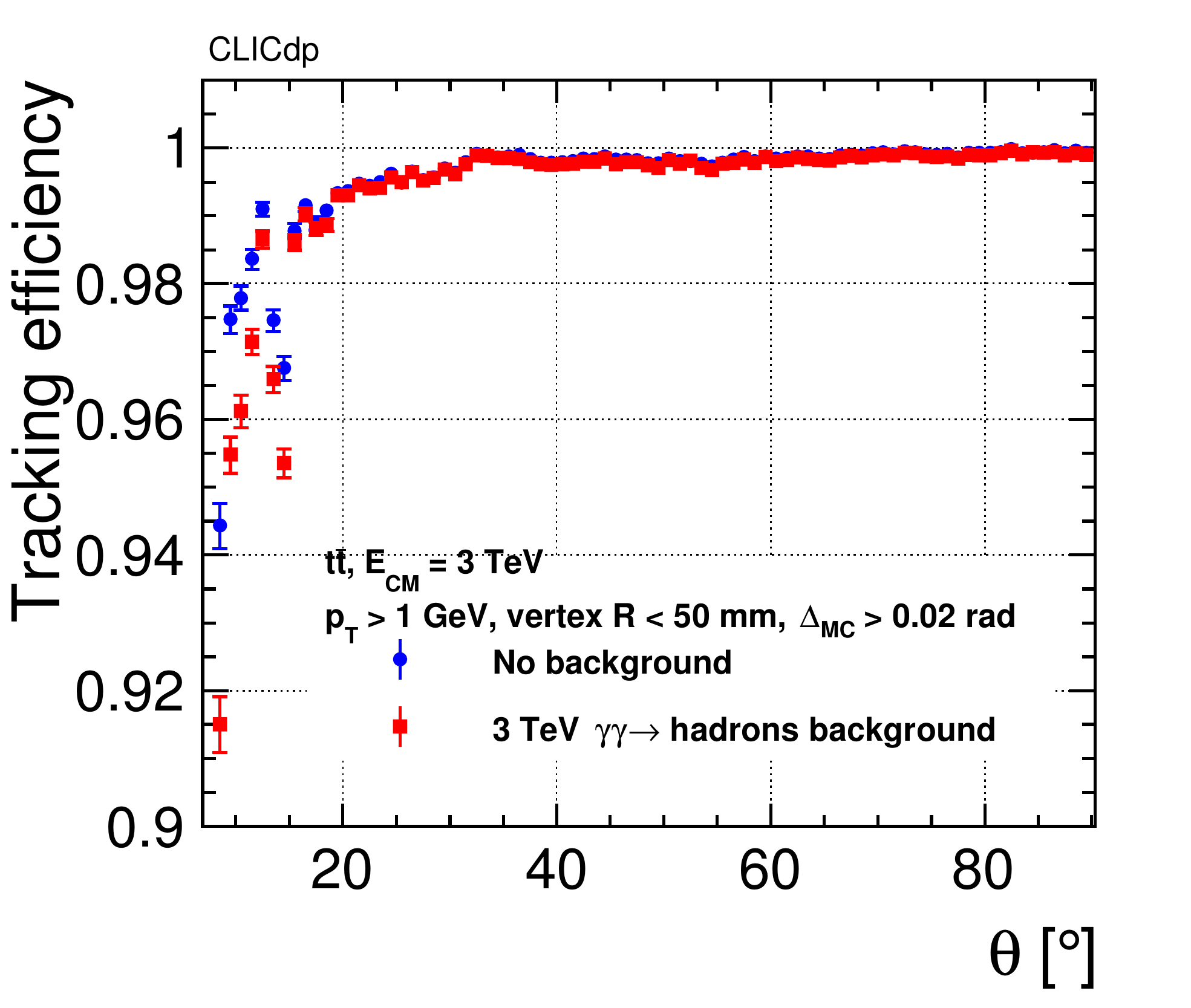}
\end{subfigure}\hfil
\begin{subfigure}{0.45\textwidth}
\includegraphics[width=\linewidth]{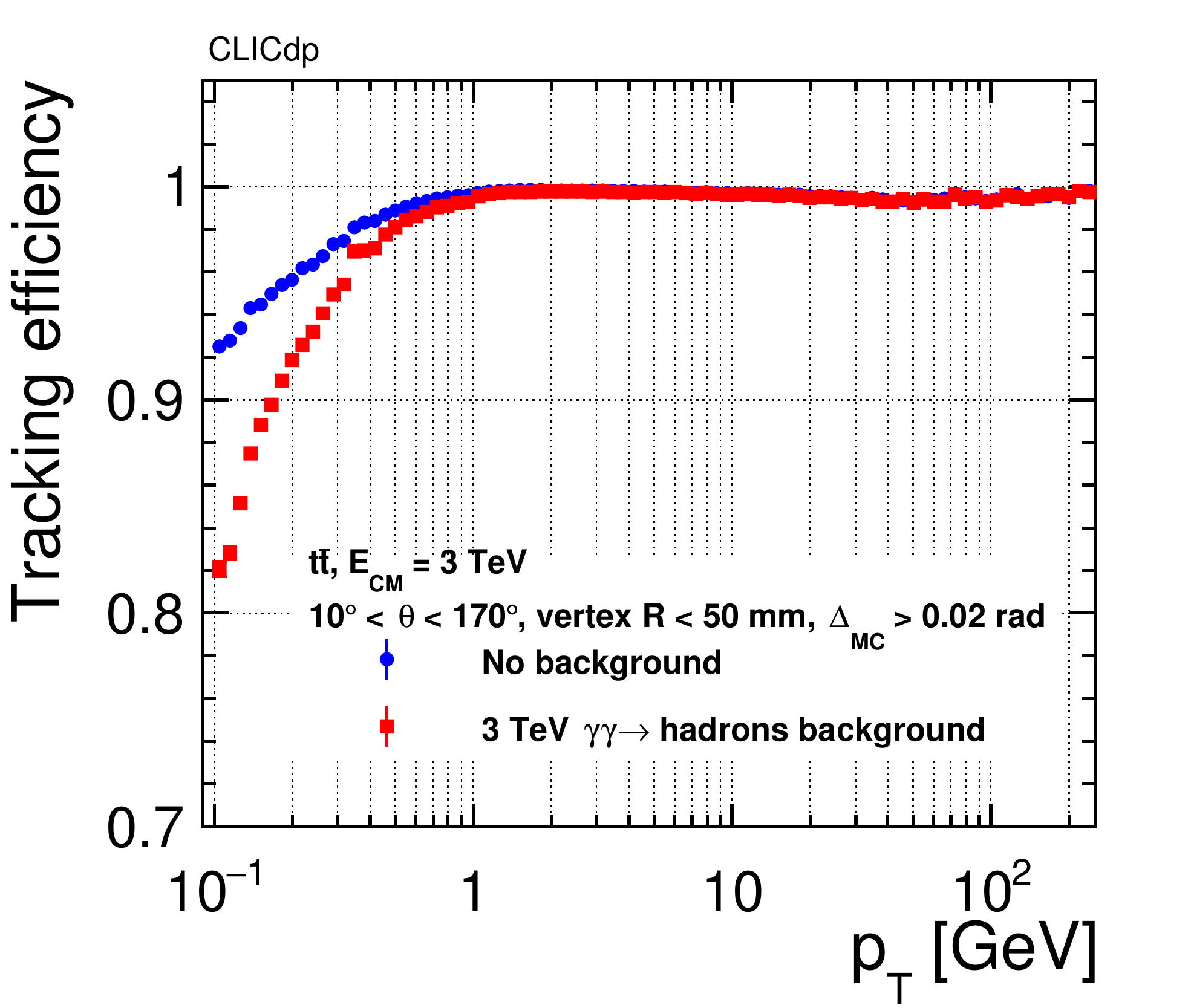}
\end{subfigure}\hfil

\medskip
\begin{subfigure}{0.45\textwidth}
\includegraphics[width=\linewidth]{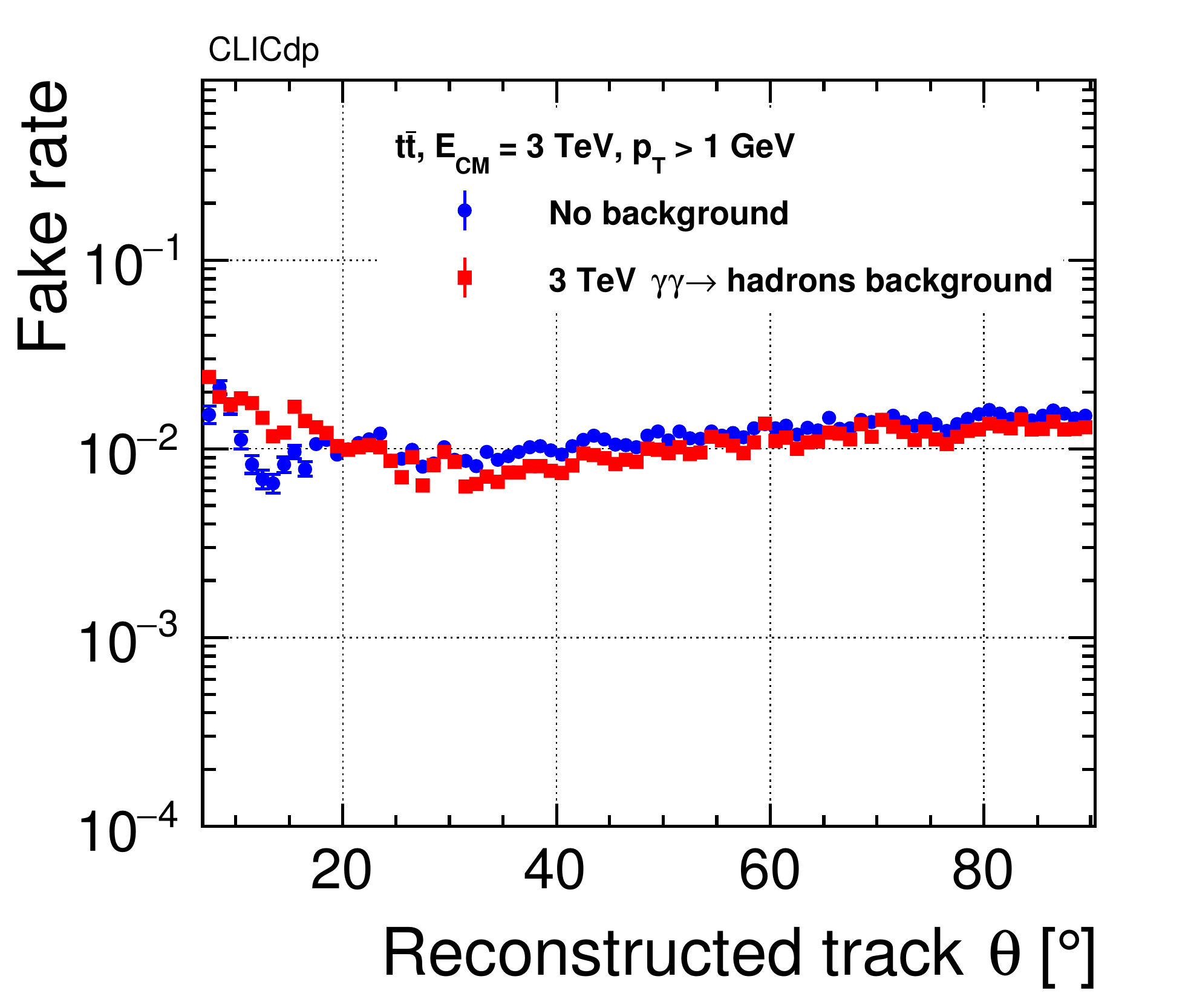}
\end{subfigure}\hfil
\begin{subfigure}{0.45\textwidth}
\includegraphics[width=\linewidth]{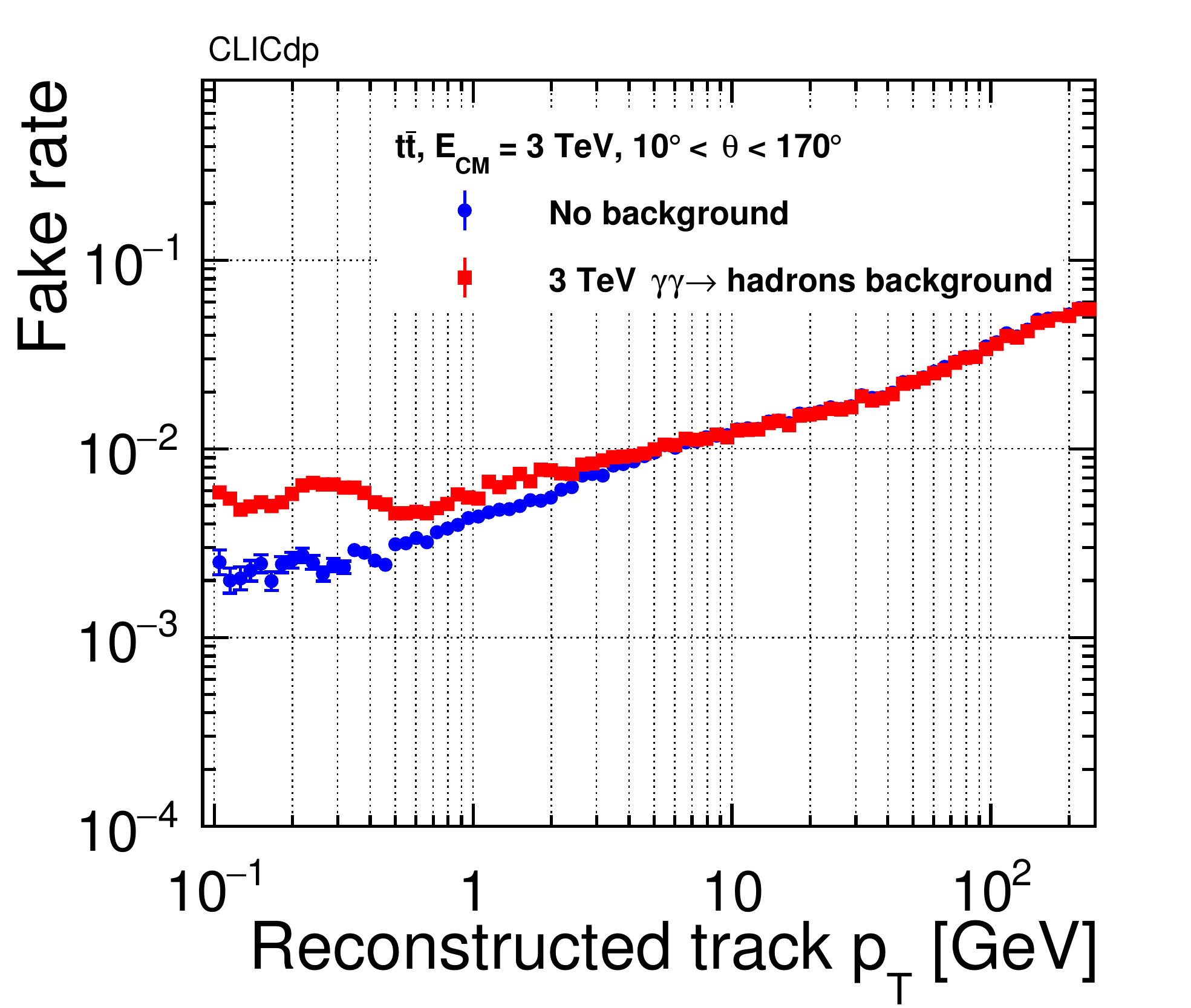}
\end{subfigure}

\caption{Tracking efficiency (top) and fake rate (bottom) as a function of polar angle (left) and \pT (right) for \epem$\rightarrow$ \ttbar events at \SI{3}{\TeV} centre-of-mass energy. Results are shown for samples reconstructed with and without the overlay of 30 bunch crossings of 
\gghad background expected at \SI{3}{\TeV} CLIC energy stage.}
\label{fig:ttbar}
\end{figure}

In the following, only the tracking performance for the highest-energy sample is discussed, as it represents the highest occupancy, thus the most challenging environment.
The efficiency and fake rate are shown as a function of polar angle and transverse momentum (\cref{fig:ttbar}). 
Low-\pT and forward region ($\theta$ < \ang{20})
constitute the most difficult area of the phase space for track reconstruction,
and therefore a loss in efficiency is expected. The fake rate is around 1\% for all reconstructed polar angles 
and increases up to 7\% in the very high-\pT region where tracks within the jet are more collimated.
Moreover, the strongest dependence on the presence of background is observed for particles with \pT smaller than \SI{1}{\GeV}. 
The tracking efficiency is otherwise close to 100\% and unaffected by \gghad, while the fake rate is increased due to the higher hit occupancy.

\begin{figure}[htb]
\centering
\includegraphics[width=0.5\linewidth]{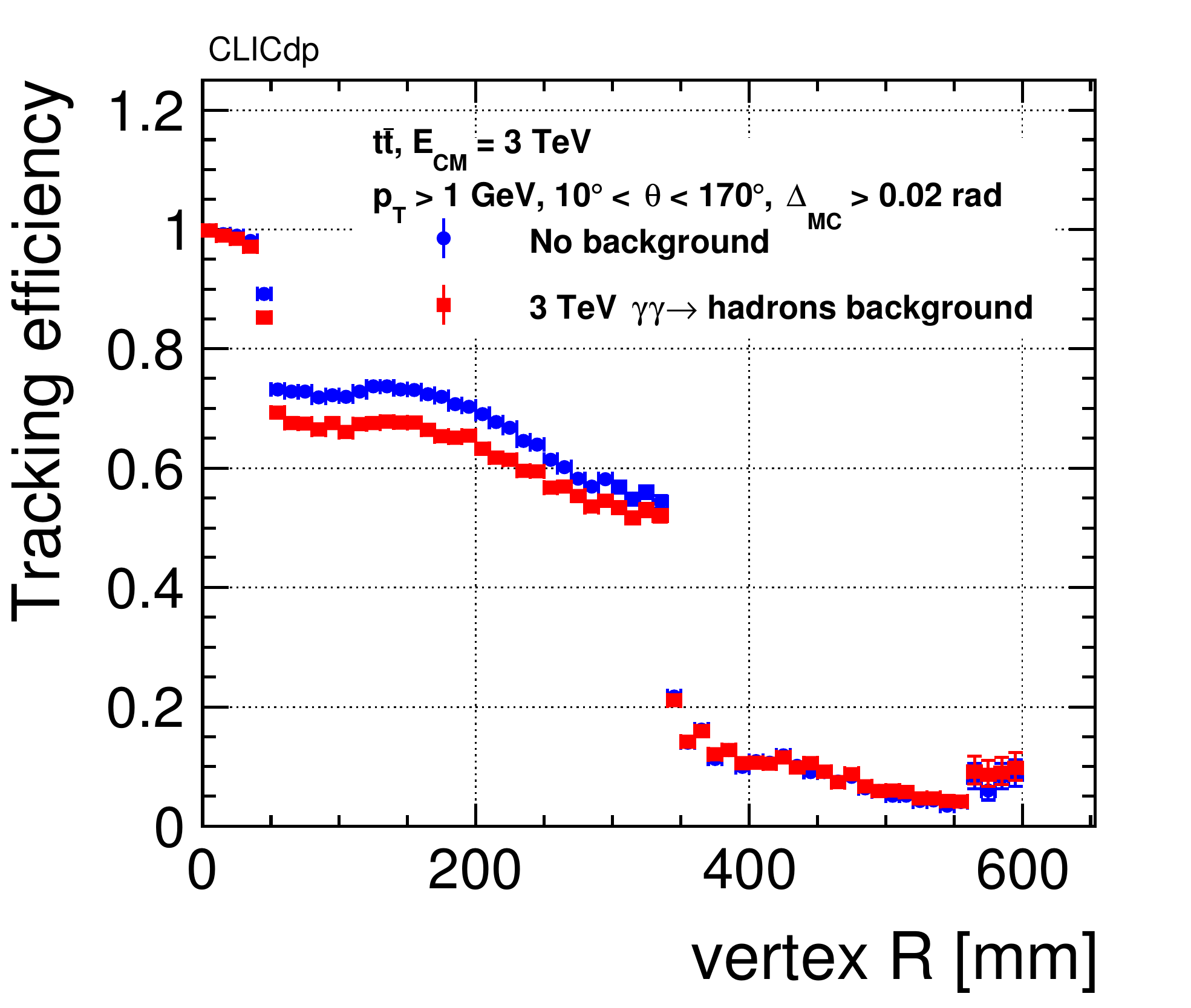}
\caption{Tracking efficiency as a function of particle production vertex radius for \epem$\rightarrow$ \ttbar events at \SI{3}{\TeV} centre-of-mass energy. Results are shown for samples reconstructed with and without the overlay of 30 bunch crossings of 
\gghad background expected at \SI{3}{\TeV} CLIC energy stage.}
\label{fig:ttbar_vtx}
\end{figure}

In the simulated \epem$\rightarrow$ \ttbar events, displaced particles are present, which may be the products of a primary particle decay, an interaction with the detector material or a radiative phenomenon. Therefore such a sample offers the possibility of scanning the tracking performance for different values of displacement. 
The tracking efficiency is studied as a function of particle production vertex radius (\cref{fig:ttbar_vtx}), which is calculated as the displacement in the transverse plane\footnote{In this case, the displacement is not only along the $y$ axis, thus it is generically computed as $\sqrt{x^{2}+y^{2}}$.}. 
Similarly as for isolated muons of \SI{1}{\GeV} momentum (\cref{fig:singlePartDisplaced}), the tracking is fully efficient as long as the particles are produced before the second double layer of the vertex detector (radius smaller than \SI{4}{\cm}). For particles produced at the outer edge of the second double layer, the efficiency starts dropping.
This is expected, as such particles leave only maximum two hits in the vertex detector, thus the extrapolation of the cellular tracks built in the tracker is more complicated and may be less accurate.
The efficiency drop observed around \SI{35}{\cm} is due to most of the particles being produced with a larger displacement than required in the CA to cross the minimum number of layers, similarly to the isolated muons produced for the study shown in~\cref{fig:singlePartDisplaced}.
The performance is in general not affected much by the presence of \gghadron background.

\subsubsection{Resolution of the track parameters}
\label{subsec:res}

The resolution of the following track parameters is presented: transverse ($d_{0}$) and longitudinal ($z_{0}$) impact parameter, azimuthal ($\phi$) and polar ($\theta$) angle, and transverse momentum.
Simulated isolated muons are used with fixed \pT and $\theta$.

The resolutions are estimated from the distribution of the track residuals, defined as the difference between the reconstructed track parameters and the simulated particle parameters.
In every bin in $\theta$ and \pT, the entire distribution is fitted with a Gaussian function, whose standard deviation defines the resolution. 
For the transverse momentum resolution, each residual is divided by the square of the estimated \pT.

\begin{figure}[!htb]
\centering

\begin{subfigure}{0.45\textwidth}
\includegraphics[width=\linewidth]{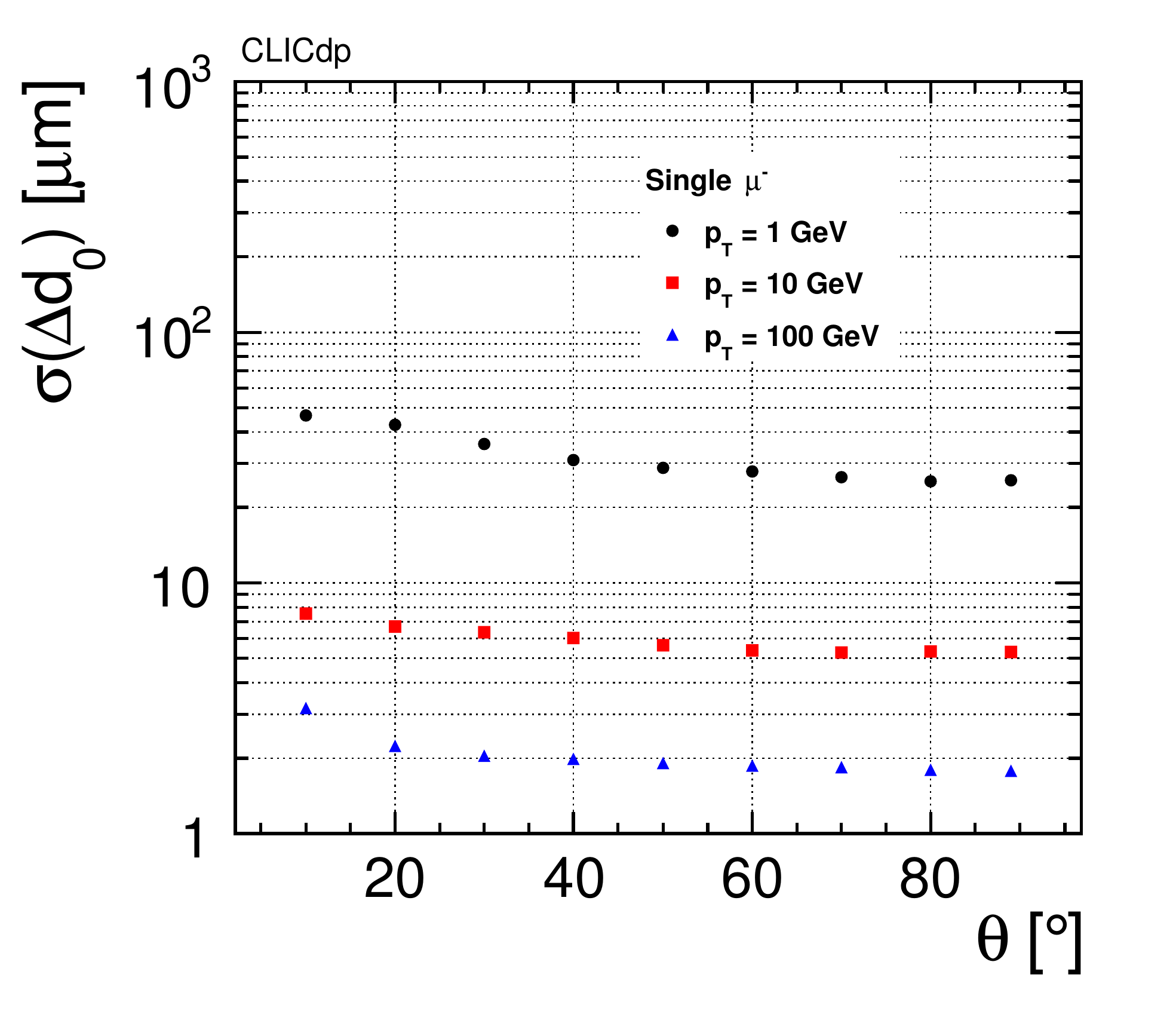}
\end{subfigure}\hfil
\begin{subfigure}{0.45\textwidth}
\includegraphics[width=\linewidth]{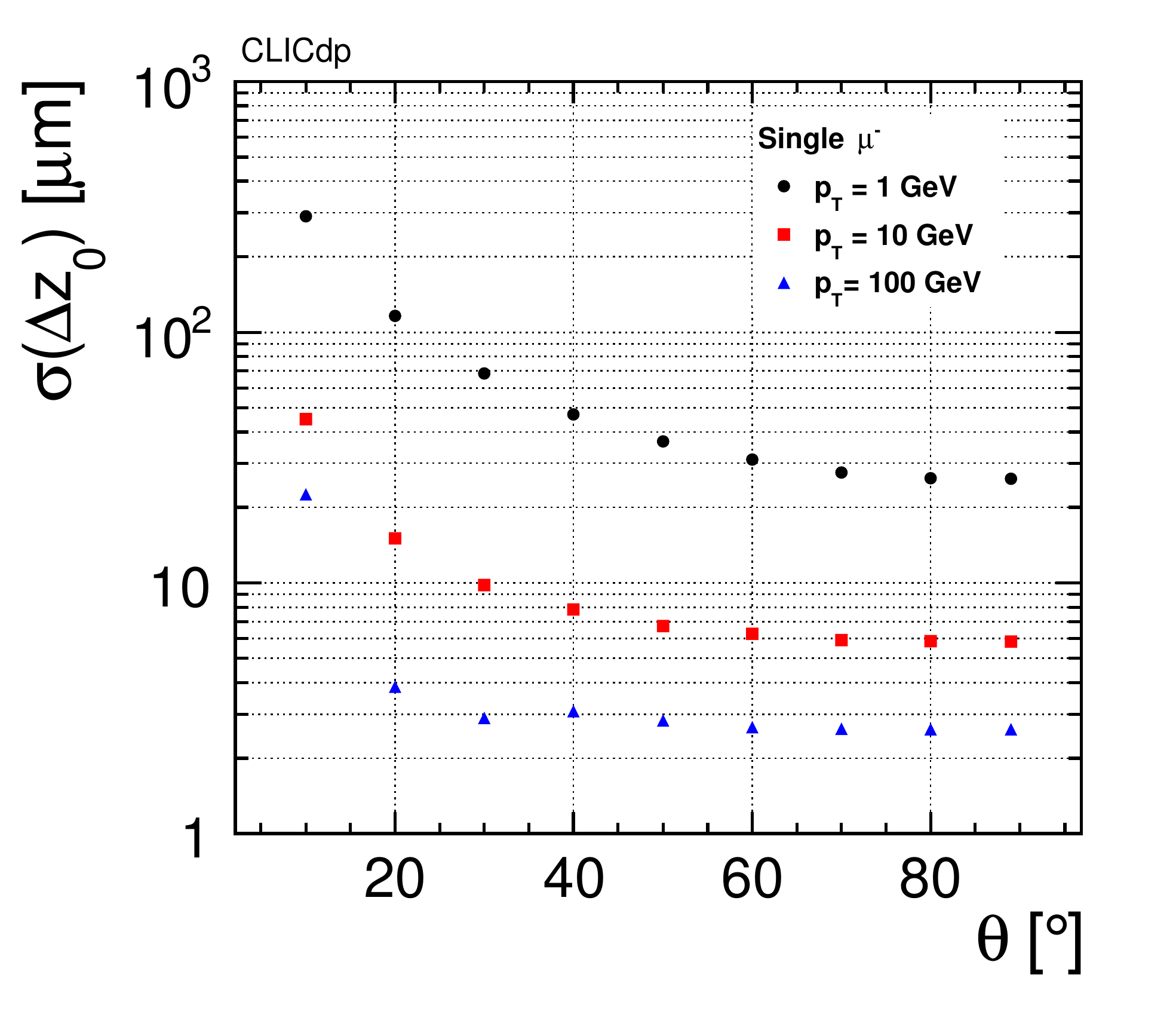}
\end{subfigure}\hfil

\medskip
\begin{subfigure}{0.45\textwidth}
\includegraphics[width=\linewidth]{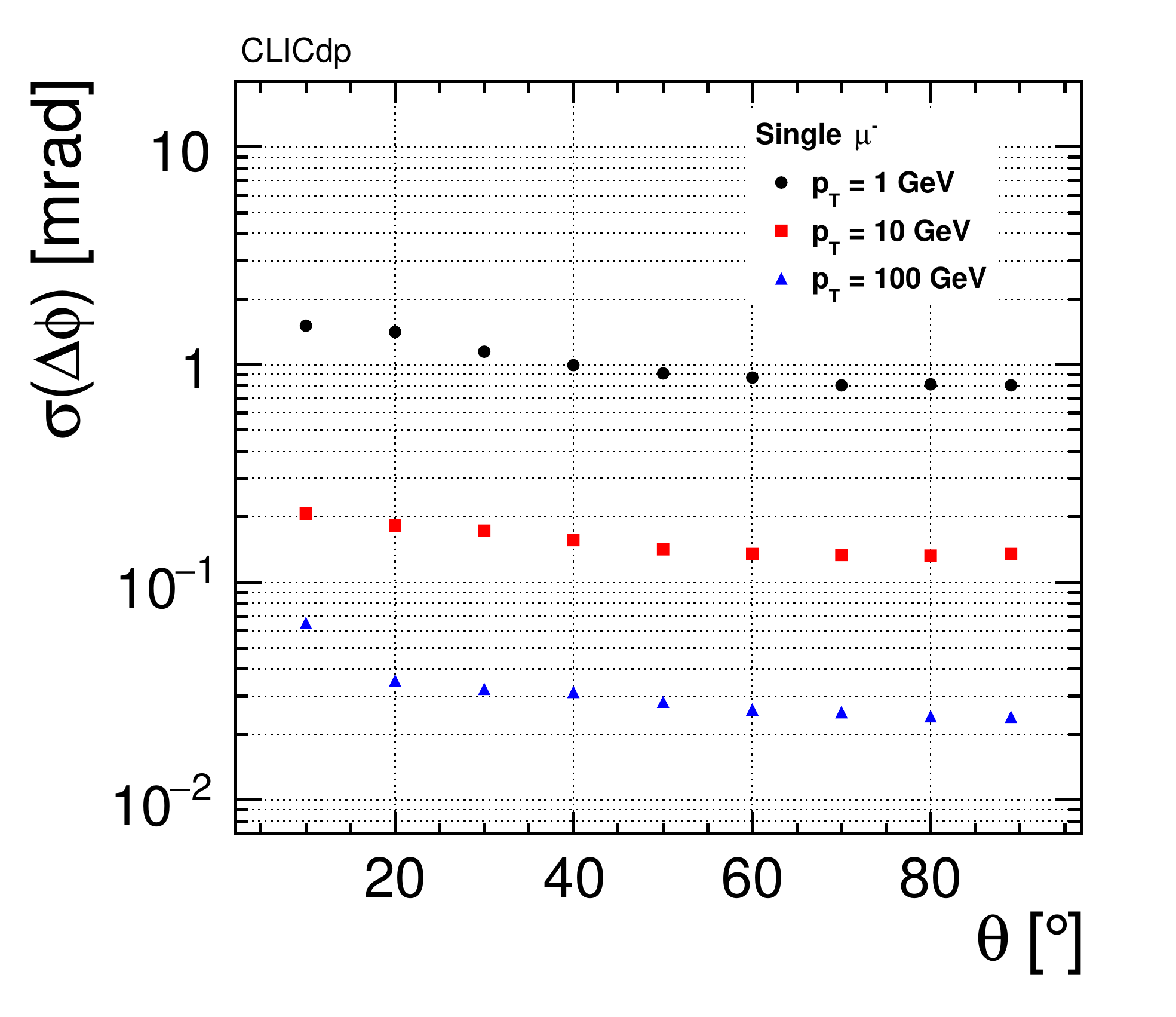}
\end{subfigure}\hfil
\begin{subfigure}{0.45\textwidth}
\includegraphics[width=\linewidth]{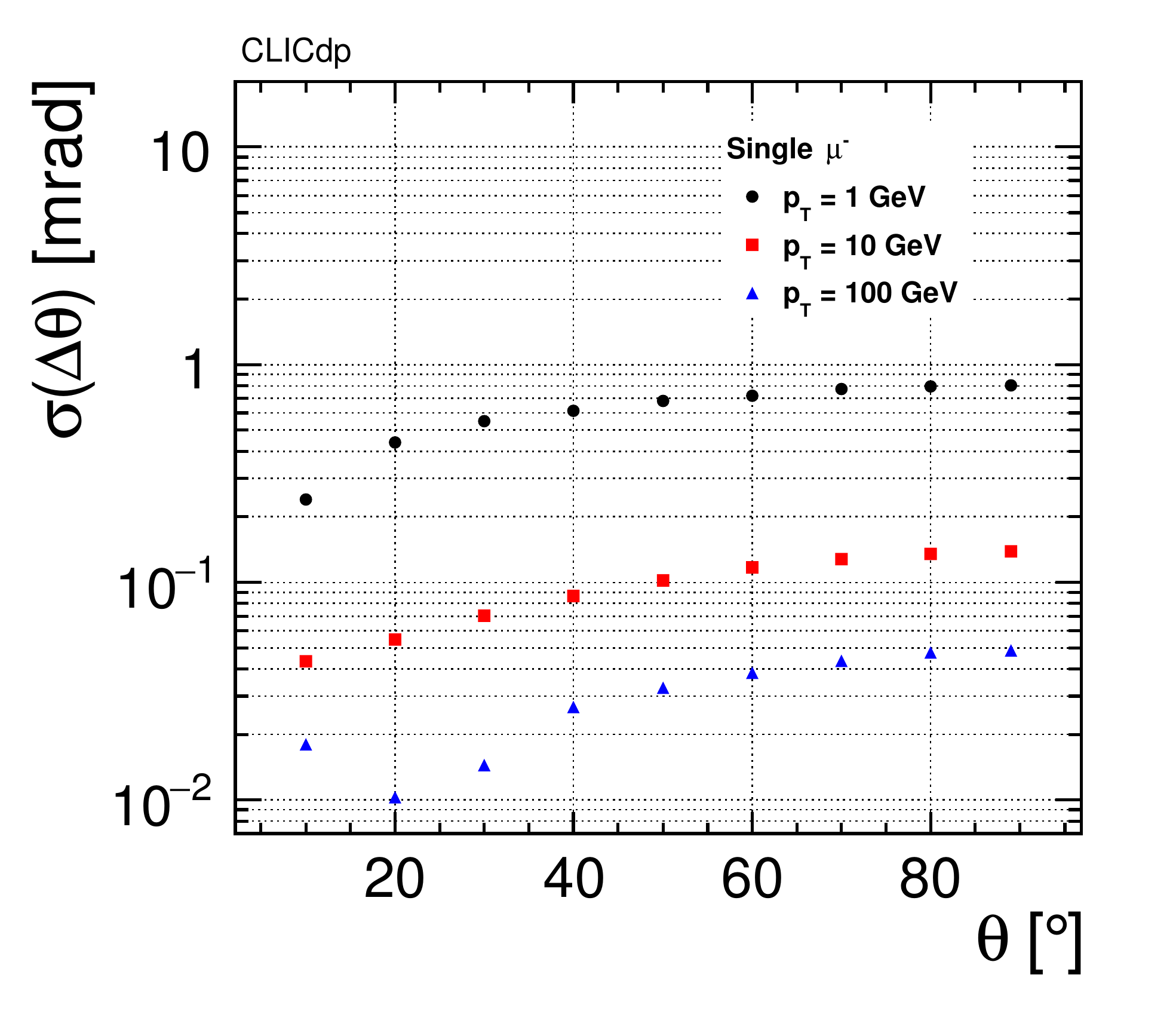}
\end{subfigure}\hfil

\medskip
\begin{subfigure}{0.45\textwidth}
\includegraphics[width=\linewidth]{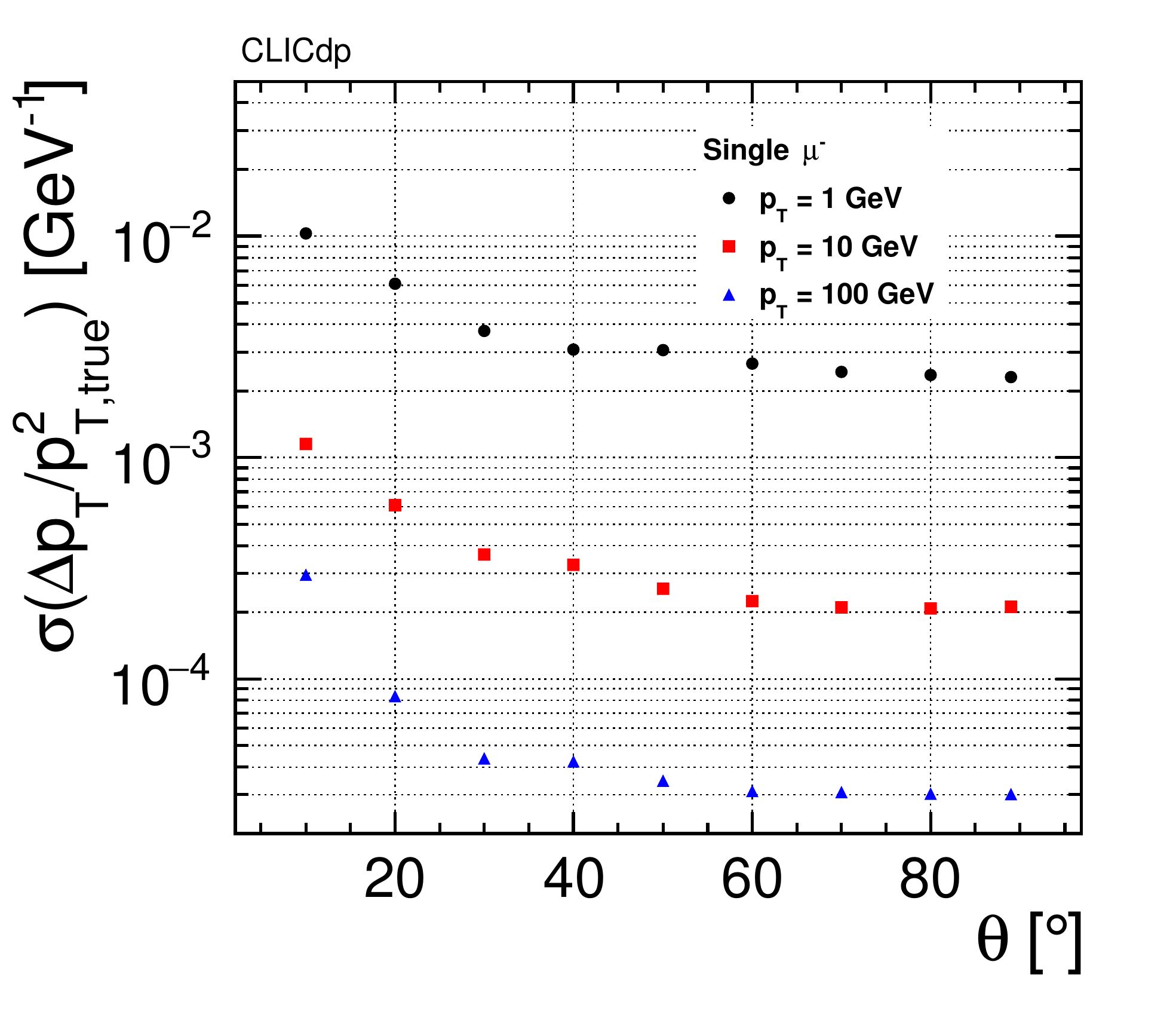}
\end{subfigure}

\caption{Resolution in track parameters for isolated muons with \pT = \SI{1}{\GeV}, \SI{10}{\GeV}, and \SI{100}{\GeV} as a function of the polar angle $\theta$. From top to bottom and left to right: transverse and longitudinal impact parameters, azimuthal and polar angle, transverse momentum.}
\label{fig:singlePart_res_theta}
\end{figure}

\begin{figure}[!htb]
\centering

\begin{subfigure}{0.45\textwidth}
\includegraphics[width=\linewidth]{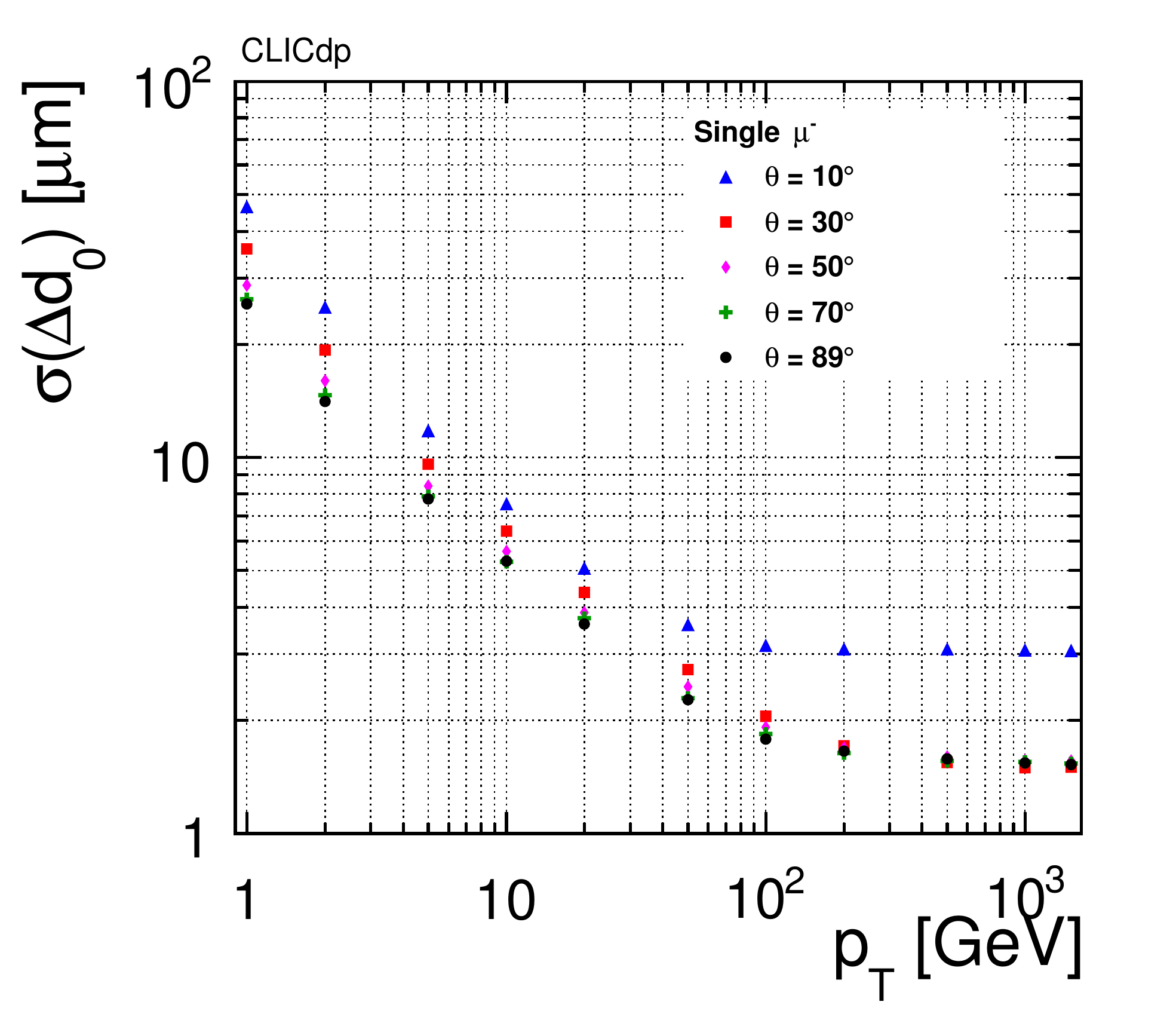}
\end{subfigure}\hfil
\begin{subfigure}{0.45\textwidth}
\includegraphics[width=\linewidth]{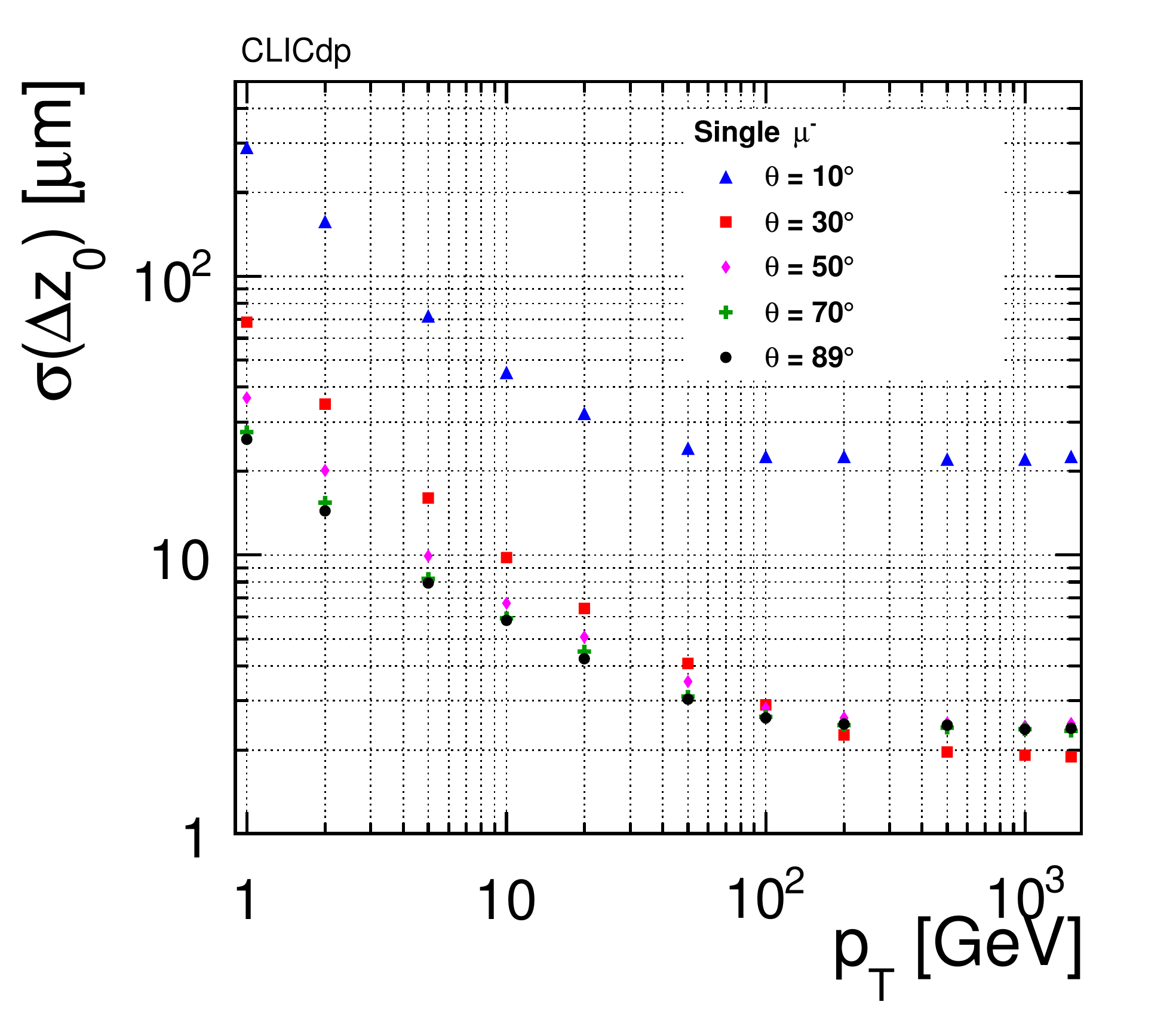}
\end{subfigure}\hfil

\medskip
\begin{subfigure}{0.45\textwidth}
\includegraphics[width=\linewidth]{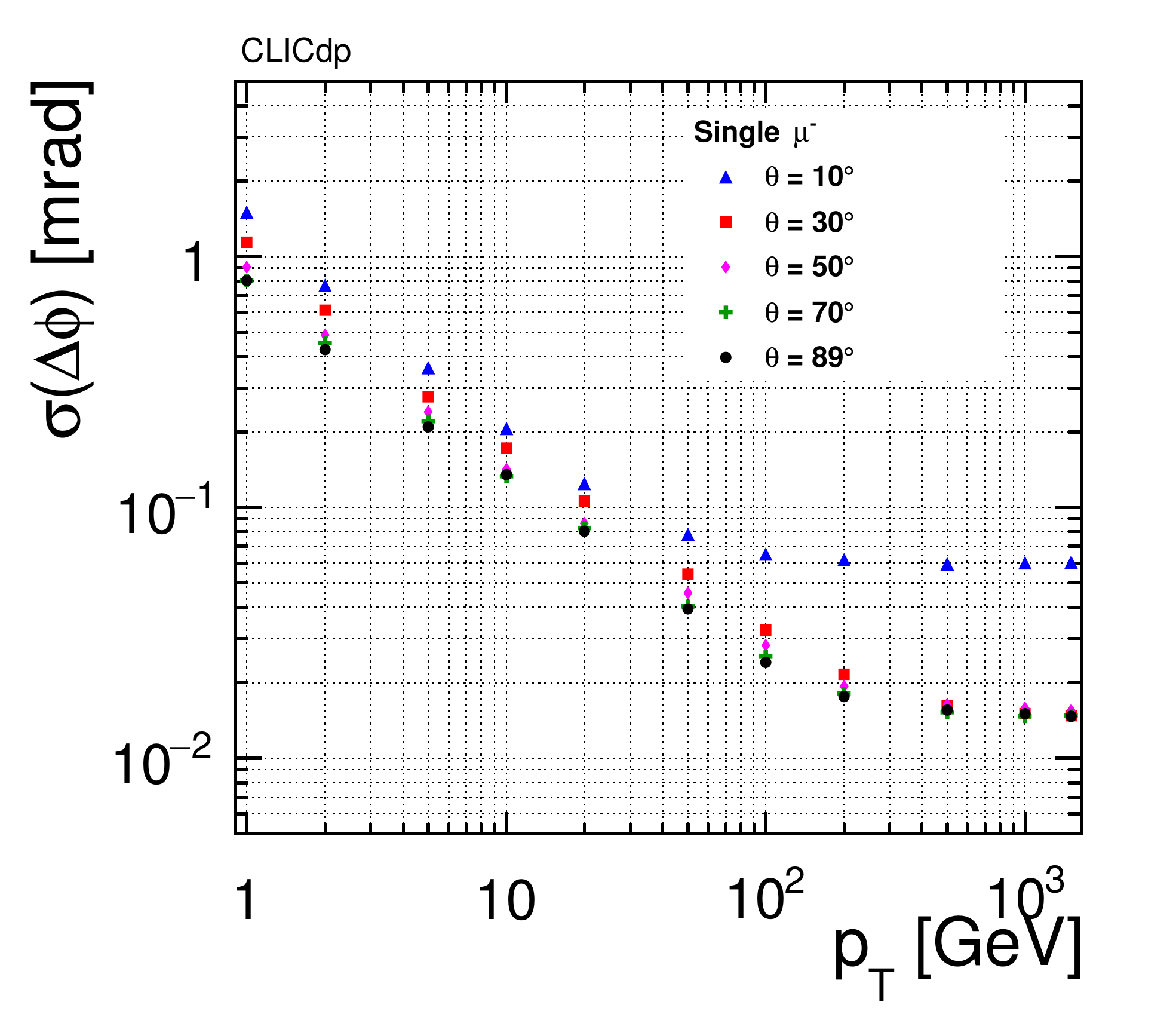}
\end{subfigure}\hfil
\begin{subfigure}{0.45\textwidth}
\includegraphics[width=\linewidth]{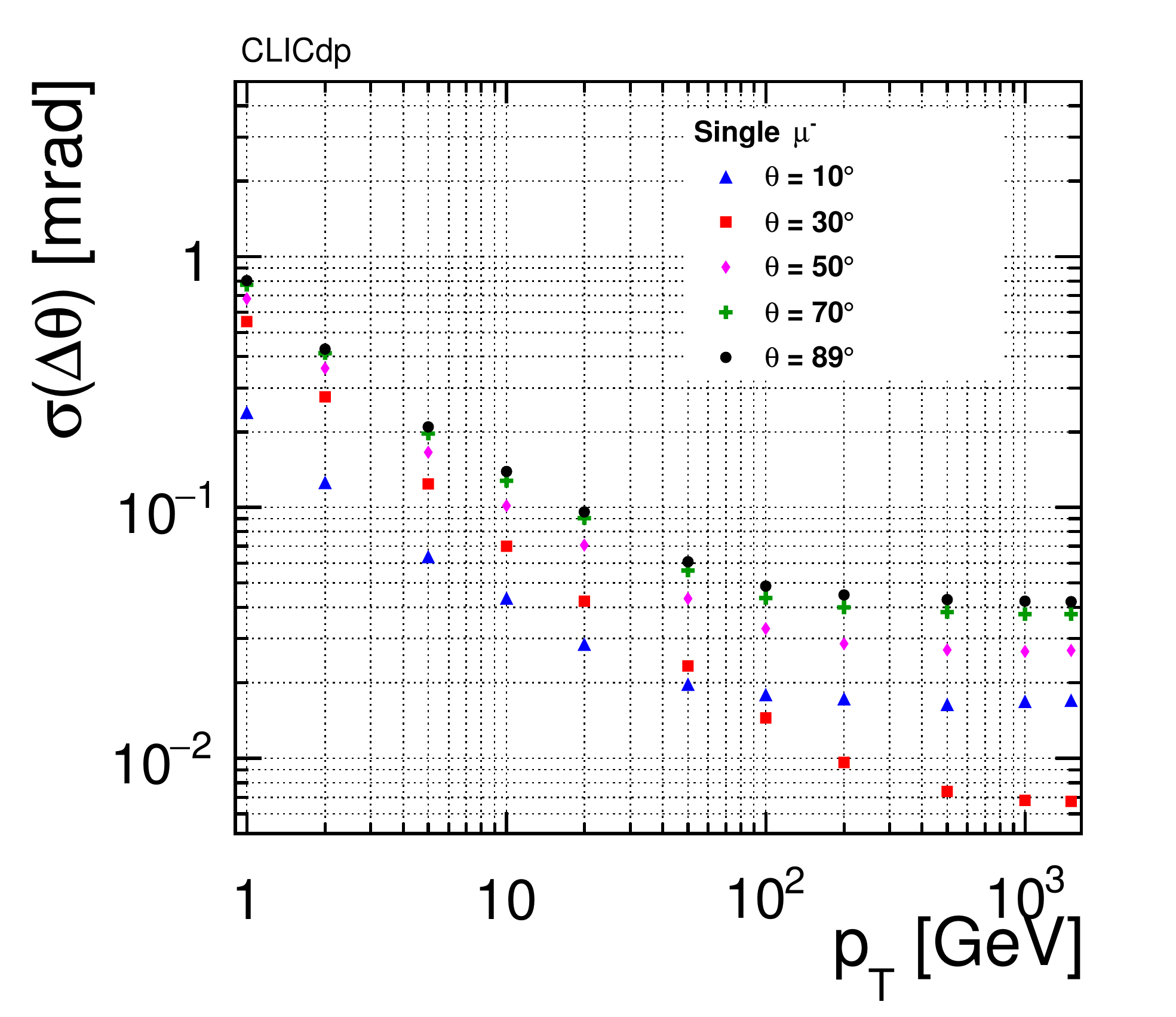}
\end{subfigure}\hfil

\medskip
\begin{subfigure}{0.45\textwidth}
\includegraphics[width=\linewidth]{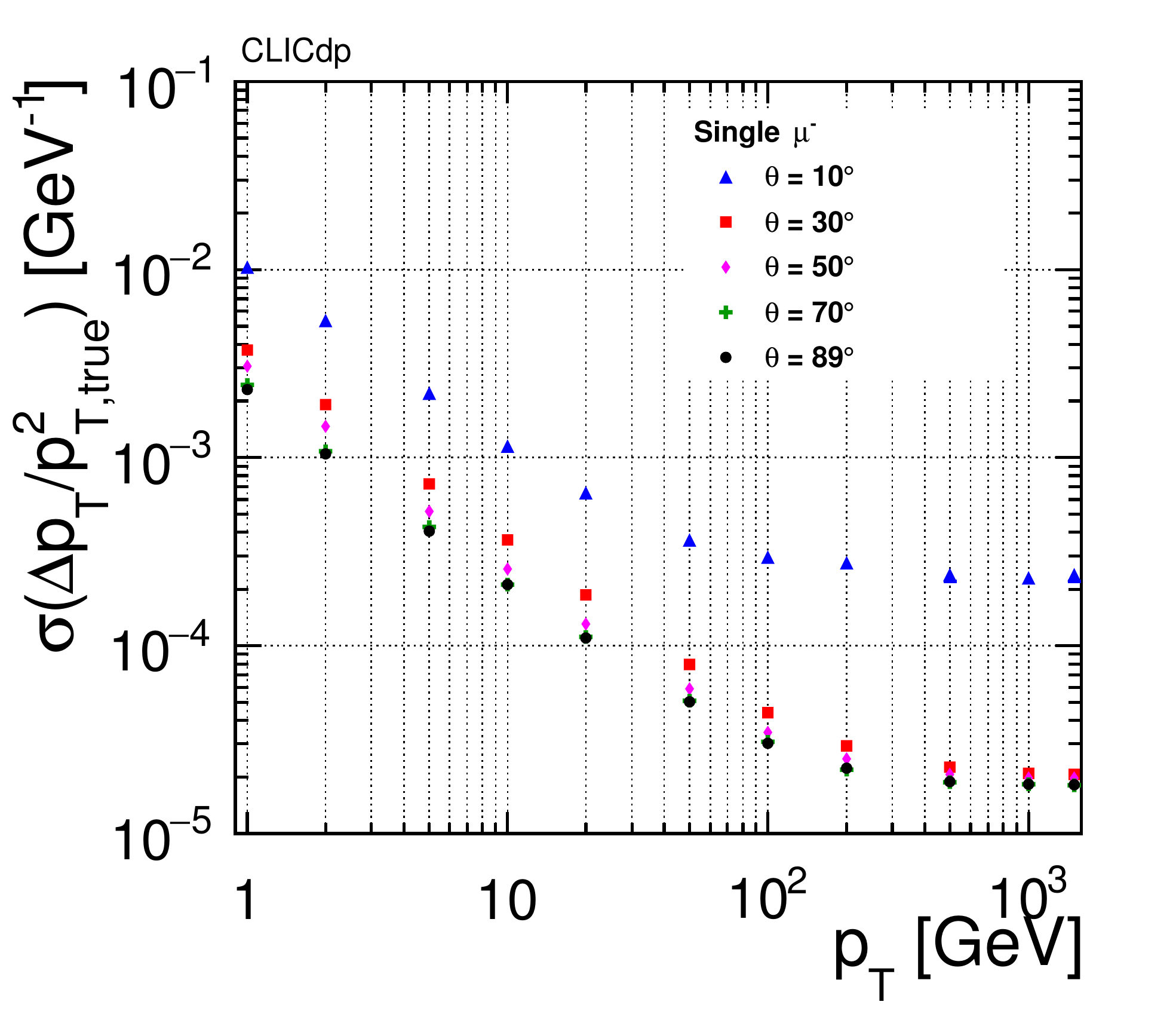}
\end{subfigure}

\caption{Resolution in track parameters for isolated muons with polar angle $\theta$ = \ang{10}, \ang{30}, \ang{50}, \ang{70} and \ang{89} as a function of \pT. From top to bottom and left to right: transverse and longitudinal impact parameters, azimuthal and polar angle, transverse momentum.}
\label{fig:singlePart_res_pt}
\end{figure}

\cref{fig:singlePart_res_theta} shows the dependence of the track parameter resolutions on polar angle $\theta$ for isolated muons with \pT = \SI{1}{\GeV}, \SI{10}{\GeV}, and \SI{100}{\GeV}. 
Track parameter resolutions are shown in \cref{fig:singlePart_res_pt} as a function of \pT for isolated muons with $\theta$ = \ang{10}, \ang{30}, \ang{50}, \ang{70} and \ang{89}. 

In general, the resolution of the impact parameters deteriorates in the most forward regions, since the distance from the innermost hit to the interaction point is larger than in the barrel (\cref{fig:singlePart_res_theta} and \cref{fig:singlePart_res_pt}, top).
Resolutions for high-momentum tracks are limited by the single point resolution in the vertex detector, while those for low-momentum tracks are dominated by multiple scattering.
Particles with \pT greater than \SI{100}{\GeV} benefit from the larger number of hits in the transition region without being affected by the multiple scattering, therefore the best resolution is observed for these high-momentum tracks at \ang{30}.
 
The resolution of the angular parameters (\cref{fig:singlePart_res_theta} and \cref{fig:singlePart_res_pt}, middle) are tightly connected with the impact parameter resolutions.
In particular, the polar angle dependences of the resolution in azimuthal angle and in $d_{0}$ are analogous, while those of the resolution in polar angle and in $z_{0}$ are inverted, since the longitudinal impact parameter is proportional to $\cot(\theta)$.
 
The resolution of the transverse momentum $\sigma$(\pT)$/$\pTsq (\cref{fig:singlePart_res_theta} and \cref{fig:singlePart_res_pt}, bottom) is affected by the length of the lever arm in the transverse plane. 
Therefore, it is worse for particles in the forward region than in the barrel.
Moreover, except for very forward tracks (\ang{10}), the transverse momentum resolution is almost independent of the polar angle for particles with \pT greater than \SI{100}{\GeV}.
In this regime, the \pT-resolution is dominated by the single point resolution in the tracker. 
This result also proves that, with the values of single point resolution for the CLICdet tracker (listed in~\cref{tab:detSize}), the physics goal of $\sigma_{\pT}/\pT^2 \leq \SI{2e-5}{\per\GeV}$ is achieved~\cref{sec:intro_CLIC}.

\subsubsection{CPU execution time}
\label{subsec:cpu}

As in many other particle physics experiments, track reconstruction is one of the most time-consuming parts of the event reconstruction.
The mean CPU time is estimated\footnote{The machine obtains a score of 25.7 with DB12~\cite{diracB12}.} 
using ten \ttbar events at \SI{3}{\TeV} centre-of-mass energy
without and with the overlay of \gghad background expected at \SI{3}{\TeV} CLIC energy stage.
If the background is not included, the mean CPU time per event is about \SI{15}{s}
for an average number of tracks of about $90$, where most of the time is spent in the fitting procedure.
If the background is included, the average number of tracks amounts to about $550$
and the mean CPU time per event is about 554~s with the CA building task being the most time consuming part.
In particular, the last CA building step (step~5) is the most expensive in terms of computing time, as can
also be expected from the looser requirements and the quadratic term included in the
mathematical approximation of the conformal track formulas.

%
%

\section{Conclusions and future developments}
\label{sec:summary}

Conformal tracking presented in this paper is a new pattern recognition technique 
for track finding, developed in the context of a low-mass detector designed for 
future electron-positron colliders.
It uses a cellular automaton-based track finding algorithm in conformal space to reconstruct prompt and non-prompt tracks.

This technique gives excellent performance in terms of efficiency and fake rate
for different single particle types as well as for complex events
and fulfills the physics requirements for track parameter resolutions.
Additionally, its robustness was proven against the challenging \gghadron beam-induced background
for the highest energy stage of CLIC.

The modularity of the software was also presented. 
Its flexible implementation allows the adaption of appropriate sets of cuts 
according to different event topologies and detector geometries.
For example, the conformal tracking has also been successfully applied
with only small parameter modifications to 
the CLIC-Like Detector (CLD) at FCC-ee, which not only has a different sub-detector design
and magnetic field magnitude, but also includes different background conditions~\cite{Leogrande:2630512,Benedikt:2651299}.
However, its mathematical foundation prevents it from being successful 
in the presence of large effects from multiple scattering; 
thus it is not a promising option for current trackers because of the larger material budget.

In the next phase, a consolidation of the conformal tracking software is planned.
Firstly, detailed studies are ongoing on its application in the case of 
small deviations from homogeneity in the detector magnetic field.
In parallel, the implementation of a robust regression technique to treat outliers within a track is being investigated.
As an example, a soft hit-to-track assignment
can replace the current strategy of refitting the track while removing hits one by one.
Moreover, the robustness of the algorithm must be maintained with the additional overlay of incoherent \epem pair events,
the second main type of background for CLIC.
Finally, the CPU performance offers room for further optimisation, 
in particular regarding the reconstruction of non-prompt and low-\pt particles. 
A first prelimiary study on reducing the number of seeds using the $z$-coordinate
information gave already about 20\% improvement.
Further improvements can be obtained also with tuning of the parameters 
and the input hit collections used in the CA building and extension steps in the final stage,
but could also be obtained in a multi-core usage mode,
taking advantage of the intrinsic parallelizable nature of the cellular automaton part of the algorithm.

Conformal tracking can also be used as fertile 
ground for more diversified tracking ideas. For example, the
algorithm could be upgraded to profit from the hit time information 
from the CLICdet tracker, currently used only in the cluster reconstruction in the calorimeters. 
This will exploit the synergies with the ongoing hardware R\&D projects 
for tracker developments both at current and future accelerators.

\section*{Acknowledgements} 
\label{sec:acknowledgements} 

The CLICdp collaboration gratefully acknowledges CERN for its continued support.
This work benefited from services provided by the ILC Virtual Organisation, supported by the national resource providers of the EGI Federation. 
This research was done using resources provided by the Open Science Grid, which is supported by the National Science Foundation and the U.S. Department of Energy's Office of Science.

\printbibliography[title=References]

\end{document}